\documentclass[pre,a4paper,10pt,twoside,showpacs,showkeys,superscriptaddress,nofootinbib,twocolumn,floatfix]{revtex4-1}
\usepackage{amsfonts,amsmath,amssymb}
\usepackage{graphicx,color}

\usepackage[bookmarks,pdfhighlight=/O,colorlinks=false,pdfstartview=FitH]{hyperref}

\newcommand{\dd}{\mathrm{d}}

\newcommand{\ee}{\ensuremath{\mathrm{e}}}
\newcommand{\ii}{\ensuremath{\mathrm{i}}}
\newcommand{\R}{\ensuremath{\mathbb{R}}}
\newcommand{\erw}[1]{\ensuremath{\left \langle {#1} \right \rangle}}
\newcommand{\kB}{\ensuremath{k_{\text{B}}}}
\DeclareMathOperator{\re}{Re}

\begin{document}

\title{Simulation of stationary Gaussian noise with regard to the
  Langevin equation with memory effect}

\author{Julian Schmidt}
\email{juschmidt@th.physik.uni-frankfurt.de}
\affiliation{Institut f{\"u}r theoretische Physik, Goethe-Universit{\"a}t Frankfurt am Main,
Max-von-Laue-Stra{\ss}e 1, 60438 Frankfurt, Germany}

\author{Alex Meistrenko}
\email{meistrenko@th.physik.uni-frankfurt.de}
\affiliation{Institut f{\"u}r theoretische Physik, Goethe-Universit{\"a}t Frankfurt am Main,
Max-von-Laue-Stra{\ss}e 1, 60438 Frankfurt, Germany}

\author{Hendrik van Hees}
\email{hees@fias.uni-frankfurt.de}
\affiliation{Institut f{\"u}r theoretische Physik, Goethe-Universit{\"a}t Frankfurt am Main,
Max-von-Laue-Stra{\ss}e 1, 60438 Frankfurt, Germany}
\affiliation{Frankfurt Institute for Advanced Studies,
  Ruth-Moufang-Stra{\ss}e 1, 60438 Frankfurt, Germany}

\author{Zhe Xu} 
\email{xuzhe@mail.tsinghua.edu.cn}
\affiliation{Department of Physics, Tsinghua University and
  Collaborative Innovation Center of Quantum Matter, Beijing 100084,
  China}

\author{Carsten Greiner}
\email{Carsten.Greiner@th.physik.uni-frankfurt.de}
\affiliation{Institut f{\"u}r theoretische Physik, Goethe-Universit{\"a}t Frankfurt am Main,
Max-von-Laue-Stra{\ss}e 1, 60438 Frankfurt, Germany}

\date{\today}

\begin{abstract}
  We present an efficient method for simulating a stationary Gaussian
  noise with an arbitrary covariance function and then study numerically
  the impact of time-correlated noise on the time evolution of a $1+1$
  dimensional generalized Langevin equation by comparing also to
  analytical results. Finally, we apply our method to the generalized
  Langevin equation with an external harmonic and double-well potential.
\end{abstract}


\maketitle

%%%%%%%%%%%%%%%%%%%%%%%%%%%%%%%%%%%%%%%%%%%%%%%%%%%%%%%%
\section{Introduction}
\label{sec:introduction}
%%%%%%%%%%%%%%%%%%%%%%%%%%%%%%%%%%%%%%%%%%%%%%%%%%%%%%%%

Brownian motion describes the rapid and irregular motion of particles in
random directions, resulting from collisions within a thermal
bath. Based on the physical motivation for the dynamics P.\ Langevin set
up a one-dimensional equation of motion which splits the force due to
the thermal bath into a macroscopic force $-\gamma \dot{x}(t)$ with
friction coefficient $\gamma$ and a microscopic fluctuating force
$\xi(t)$,
\begin{equation}
m\ddot{x}(t)+\gamma \dot{x}(t) = \xi(t)\,.
\end{equation}
This stochastic equation is the original Langevin equation, where
$\xi(t)$ stands for a stochastic process with a vanishing expectation
value, since there is no preferred direction for the
collisions. According to the stochastic nature of $\xi(t)$, it is also
called noise. In the original Langevin equation the noise term is
$\delta$-correlated and called white noise. ``\textit{Although pure
  white noise does not occur as a physically realizable process}'', it
has been studied intensively ``\textit{as an idealization of many real
  physical processes}'' \cite[p.~63]{TLE}. When the noise term is no
longer $\delta$-correlated it is called colored noise, leading to a
non-Markovian random process with memory effects in the corresponding
generalized Langevin equation.

The generalized Langevin equation has been applied to a wide range of
physical topics: In ultra-relativistic heavy ion collisions, for
instance, disoriented chiral condensates \cite{XU1,Xu:1999aq} and the
effects of dissipation in the deconfining \cite{FRAGA2006} and the
chiral \cite{Nahrgang:2011mv,Herold:2013bi} phase transitions have been
investigated. It has also found applications in realistic
field-theoretical descriptions of the dynamics of phase transitions
\cite{CALCLA, GLEISER1993, RISCHKE1998,
  FARIAS2008,Farias:2009stochastic}, semiclassical approximations for
the dynamics of quantum fields \cite{Greiner:1996dx}, the interpretation
of the Kadanoff-Baym equations in non-equilibrium quantum field theory
\cite{Greiner:1998vd} as well as in condensed matter physics, e.g., in
the characterization of heat conduction in low-dimensional systems
\cite{DHAR2008} or in order to model molecular dynamics, as for example
at molecular junctions \cite{LUE2012} or reaction-rate theory
\cite{Haenggi:1990rmp}. In astronomy the motion of accretion disks
around compact astrophysical objects have recently been studied under
the model assumption of a generalized Langevin equation
\cite{HARKO2014}. In biology the fluctuations within single protein
molecules can also be described by generalized Langevin equations
\cite{MIN2005}. This list only gives a few examples and is far from
being complete. The effects of a non-Markovian dissipation kernel and
colored noise in the context of quantum-Brownian motion have been
studied in \cite{Fraga:2014kva}. The importance of the implementation of
memory effects and colored noise to describe causal baryon diffusion to
describe the relativistic motion of the hot and dense matter created in
heavy-ion collisions has been emphasized in \cite{Kapusta:2014dja}.

With this motivation for the applications of non-Markovian Langevin
dynamics with colored noise we show in Section~\ref{sec:noise} how
stochastic processes with stationary Gaussian noise can be defined and
effectively simulated for any given covariance function. The
time-correlated noise leads to interesting memory effects in the
numerical solution of the generalized Langevin equation, derived in
Section \ref{sec:langevinColour}. As first feasibility tests of our
method we consider the generalized Langevin equation for different
classical-mechanics setups: particles without an external potential
(Section~\ref{sec:noPotential}), with a harmonic
(Section~\ref{sec:harmonic}) as well as a double-well potential,
including the symmetric (Section~\ref{sec:langevinDoubleWellSym}) and
the asymmetric cases (Section~\ref{sec:langevinDoubleWellAsym}). While
for the free particle and the particle in a harmonic potential analytic
solutions are available to validate our numerical method, for the
double-well potentials, only numerical results are presented.

%%%%%%%%%%%%%%%%%%%%%%%%%%%%%%%%%%%%%%%%%%%%%%%%%%%%%%%%
\section{Derivation of a general stationary Gaussian process}
\label{sec:noise}
%%%%%%%%%%%%%%%%%%%%%%%%%%%%%%%%%%%%%%%%%%%%%%%%%%%%%%%%

A Gaussian process can be described by its expectation value and its
covariance function. We present a method to generate a stationary
Gaussian process for an arbitrary covariance function.  The sum
\begin{equation}
\label{eq:generalNoise}
\xi(t)=\sum_{i=1}^n a_i b(t-t_i),\quad t\in[0,T]
\end{equation}
with a stochastic amplitude $a_i$ describes a very general stochastic
process with $n$ discrete pulses at times $t_1,... ,t_n$ in the observed
time interval $[0,T]$ and $b(t)$ denoting an arbitrary pulse shape
\cite[p.~419]{HEER}. The noise shall have the following attributes:
\begin{enumerate}
\item The expectation value of the noise vanishes,
\[\langle \xi(t) \rangle=0 \Leftrightarrow \langle a_i(t) \rangle=0 \,.\]
\item The exact knowledge of the probability density of $p(a_i)$ is of
  no importance. Its characteristic function is
\begin{equation}
\label{eq:charHeight}
W(\omega) = \int_{-\infty}^{\infty}\!p(a_i) \ee^{\ii\omega a_i} \mathrm{d}a_i \,.
\end{equation}
\item The probability density $p_{t_i}$ of having a pulse at a certain
  time $t_i$ is equal to the probability density $p_{t_j}$ at a
  different time $t_j$. So for one pulse in the time interval $[0,T]$
  the probability density is
\begin{equation*}
p_{t_i}=\frac{1}{T} \,.\\
\end{equation*}
\item The probability that $n$ independent pulses occur during the time
  interval shall be given by the Poisson distribution
  \cite[p.~420]{HEER}
\begin{equation}
\label{eq:poisson}
P_n=\frac{\bar{n}^n}{n!}\ee^{-\bar{n}} \,.
\end{equation}
Here $\bar{n}$ denotes the mean number of pulses in the time interval
$[0,T]$ and can also be written as $\bar{n}=\mu T$, where $\mu$ is the
mean pulse rate.
\end{enumerate}
The total probability density for the occurrence of $n$ pulses with a
pulse height $a_i$ at times $t_0 \ldots t_n$ can be expressed as
\begin{equation}
\begin{split}
P_n[\xi(t)] &= P_n \, p(a_1) \cdots p(a_n) p_{t_1} \cdots p_{t_n} \\
			&= P_n \, p(a_1) \cdots p(a_n)T^{-n} \,.
\end{split}
\end{equation}
The path integral for fixed $n$ is the integration along all possible
times $t_1 \ldots t_n$ in the interval $[0,T]$ and all possible pulse
heights $a_1 \ldots a_n$:
\begin{equation*}
\mathrm{D}_n[\xi(t)] = \mathrm{d}t_1 \ldots \mathrm{d}t_n \, \mathrm{d}a_1 \ldots \mathrm{d}a_n \,,
\end{equation*}
resulting in
\begin{equation*}
P[\xi(t)]\mathrm{D}[\xi(t)] = \sum_{n=0}^\infty P_n[\xi(t)]\mathrm{D}_n[\xi(t)]\,.
\end{equation*}
This leads to the characteristic functional of the stochastic process $\xi(t)$ with an arbitrary auxiliary test function $k(t)$:
\begin{equation}
\begin{split}
  \Phi[k(t)]&=\int\!\exp\left[\ii \int_{-\infty}^{\infty}\!\mathrm{d}t\,
    k(t)\xi(t)\right]P[\xi(t)]\,\mathrm{D}[\xi(t)]\\
  &= \sum_{n=0}^{\infty}P_n \prod_{j=1}^n\int_0^T \!
  \frac{\mathrm{d}t_j}{T}\int_{-\infty}^{\infty}\!\mathrm{d}a_j\,p(a_j)\\
  &\phantom{=\sum_{n=0}^{\infty}\;} \times \exp\left[i a_j
    \int_{-\infty}^{\infty}\!\mathrm{d}t\, k(t)b(t-t_j)\right] \,.
\end{split}
\label{eq:charFunc}
\end{equation}
Using the characteristic function $W(\omega)$ of $p(a_i)$ (see
Eq.~\eqref{eq:charHeight}) and the Poisson distribution $P_n$
Eq.~\eqref{eq:poisson} with the mean pulse rate $\mu=\bar{n}/T$,
relation \eqref{eq:charFunc} can be transformed to
\begin{align*}
\Phi&[k(t)] = \\
&\exp\left[-\mu \int_0^T \!\mathrm{d}s \left\lbrace 1 -  \, W \! \left(\int_{-\infty}^{\infty}\!\mathrm{d}t\, k(t)b(t-s)\right) \right\rbrace \right]\,.
\end{align*}

If we now use the Taylor expansion of $W(\omega)$, we obtain
\begin{equation}
\begin{split}
W(\omega)&=1+\ii \omega\langle a_i \rangle-\frac{1}{2!}{\omega}^2\langle a_i^2 \rangle + \cdots \\
		&=1-\frac{1}{2}\sigma^2\omega^2 + \cdots
\end{split}
\end{equation}
with $\langle a_i \rangle=0$ and $\sigma^2$ as variance of $p(a_i)$. The
characteristic function now reads
\begin{align*}
\Phi[k(t)] = \exp&\bigg[-\frac{\mu \sigma^2}{2}\int_0^T \!\mathrm{d}s \int_{-\infty}^{\infty}\!\mathrm{d}t\\
&\int_{-\infty}^{\infty}\!\mathrm{d}t'\,k(t)k(t')b(t-s)b(t'-s) +\ldots \bigg] \,.
\end{align*}
For the limit $\sigma \rightarrow 0, \quad \mu \rightarrow \infty, \quad
\mu\sigma^2=\mathrm{const}$, the additional terms vanish and the
characteristic function has the form of a Gaussian process with
vanishing expectation value:
\begin{equation}
\Phi_{\mathrm{Gauss}}[k(t)] =
\exp\left[-\frac{1}{2}\int \mathrm{d}t \int \mathrm{d}t'\,
  k(t)k(t')C(t,t')\right]
\end{equation}
where
\begin{equation}
\label{eq:cor1}
C(t,t') = \left\langle\xi(t)\xi(t') \right\rangle = \mu \sigma^2
\int_0^T \!\mathrm{d}s\,b(t-s)b(t'-s) \,.
\end{equation}
By definition the process is called stationary, if $C(t,t')=C(t-t')$ for
all $t,t' \in [0,T]$. We have shown that in the limit of a small
variance $\sigma^2$ and large pulse rate $\mu$ our general noise
function (\ref{eq:generalNoise}) becomes a Gaussian process and is
therefore called Gaussian noise. Apart from knowing the variance
$\sigma^2$ of the probability distribution $p(a_i)$, we do not need any
further knowledge about that function. For simplicity, we choose a
Gaussian distribution. For $b(t)=\sqrt\frac{D}{\mu\sigma^2}\delta(t)$ we
obtain $\delta$-correlated white noise with a positive value $D$,
\begin{equation}
\xi_w(t) = \sqrt{D} \bar{\xi}_w(t)\,,
\end{equation}
where
\begin{equation}
\label{eq:normWhiteNoise}
\bar{\xi}_w(t) = \sum_{i=1}^n \frac{a_i}{\sigma} \frac{1}{\sqrt{\mu}}
\delta(t-t_i)= \sum_{i=1}^n \frac{\bar{a}_i}{\sqrt{\mu}} \delta(t-t_i)
\end{equation}
is normalized white noise with $\bar{a}_i:=\frac{a_i}{\sigma}$ being a
random Gaussian distributed variable, scaled to the variance of unity.

For a large pulse rate $\mu$ the relative variance $\frac{\Delta
  n}{\bar{n}}$ of the number of pulses becomes small. For this reason we
can fix the number of pulses in the time interval $[0,T]$ to $n$ and
split the time interval into $n$ time steps $t_i$ with step width
$\Delta t=t_{i+1}-t_i$. The pulse rate density is then given by
$\mu=\frac{1}{\Delta t}$ resulting in the following approximation for
the $\delta$-function:
\begin{equation}
  \delta(t)=\begin{cases}
    \frac{1}{\Delta t} \quad &\mathrm{for} \; t=0\\
    0 \quad &\mathrm{for} \; t \neq 0
\end{cases}.
\end{equation}
This leads to
\begin{equation}
\xi_w(t_i) = \sqrt{D} \frac{\bar{a}_i}{\sqrt{\Delta t}} \,.
\end{equation}
We return to the general expression of the covariance function as given
in Eq.\ (\ref{eq:cor1}). Since we want the process to be stationary, the
following assumptions for the pulse shape $b(t)$ are needed:
\begin{enumerate}
\item $b(t)$ is symmetric with respect to the origin $t=0$,
\begin{equation*}
  b(t) = b(-t) \,  .
\end{equation*}
\item $b(t)=0$ outside a defined interval $[-\Delta, \Delta]$.
\end{enumerate}
Substituting $t'':=-t+s$ and using the first assumption results in
\begin{equation*}
C(t,t') = \mu \sigma^2 \int_{-t}^{T-t} \!\mathrm{d}t'' \,b(t'')b(t''+t-t') \,.
\end{equation*}
The second assumption leads to a stationary stochastic process for the
interval
\begin{equation}
\label{eq:borderCondition}
t,\,t'\in [\Delta, T-\Delta] \,,
\end{equation}
because boundary effects of the interval $[0,T]$ need to be
excluded. With $t$ and $t'$ restricted to that interval, we can expand
the integration for $t''$ to infinity since $b(t)$ vanishes outside the
interval $[-\Delta,\Delta]$:
\begin{equation}
\begin{split}
\label{eq:covarianceNoise1}
C(t,t') &= \mu \sigma^2 \int_{-\infty}^{\infty}
\!\mathrm{d}t''\,b(t'')b(t''+t-t') \\
		&= \mu \sigma^2 \int_{-\infty}^{\infty} \!\mathrm{d}t''\,b(t'')b(t''+|t-t'|)\\
		&= C(|t-t'|) \,.
\end{split}
\end{equation}
This shows that the process indeed becomes stationary under the
assumptions 1, 2 and the condition~\eqref{eq:borderCondition}. From the
Wiener-Khinchin theorem we obtain the spectral density of a stationary
process as the Fourier transform of the covariance function,
\begin{equation}
\begin{split}
\label{eq:covarianceNoise2}
S_{\xi}(\omega) &= \mathcal{F}[C](\omega) \\
&= \mu \sigma^2 \mathcal{F} \left[\int_{-\infty}^{\infty}
  \!\mathrm{d}t''\,b(t'')b(t''+t))\right]\\
&= \mu \sigma^2 |\tilde{b}(\omega)|^2 \,.
\end{split}
\end{equation}
Thereby we adopt the following convention for the Fourier transform and
its inverse:
\begin{equation}
\begin{split}
  & \mathcal{F}[f](\omega) =\tilde{f}(\omega)=\int_{-\infty}^{\infty}
  \dd t \,
  f(t) \exp(\ii \omega t), \\
  & \mathcal{F}^{-1}[\tilde{f}](t) = \int_{-\infty}^{\infty} \frac{\dd
    \omega}{2 \pi} \tilde{f}(\omega) \exp(-\ii \omega t).
\end{split}
\end{equation}
Eq.~\eqref{eq:covarianceNoise2} contains the same information as
Eq.~\eqref{eq:covarianceNoise1} due to the properties of the Fourier
transform, which is independent of the sign of
$\tilde{b}(\omega)$. Therefore we chose $\tilde{b}(\omega)$ as a real
positive valued function: $b(\omega)\geq 0$ for all $\omega \in
\R$. This implies the following relations for the pulse shape,
\begin{equation}
  \begin{split}
\label{eq:covarianceNoise3}
\tilde{b}(\omega) = & \frac{1}{\sigma\sqrt{\mu}} \sqrt{S_{\xi}(\omega)} \,,\\
b(t) = &\mathcal{F}^{-1}\left[\tilde{b}\right](t) :=
\frac{1}{\sigma\sqrt{\mu}} G(t) \,,
\end{split}
\end{equation}
where 
\begin{equation}
  \quad G(t):= \mathcal{F}^{-1} \left[\sqrt{S_{\xi}}\right](t)\,,
\end{equation}
leading to $b(t)=0 \Leftrightarrow G(t)=0$. Hence the interval
$[-\Delta, \Delta]$ can be defined as the range, where $G(t) >
0$. Numerically we introduce a cut-off scale, such that $G(t)$ drops to
a sufficiently small value.

%%%%%%%%%%%%%%%%%%%%%%%%%%%%%%%%%%%%%%%%%%%%%%%%%%%%%%%%%%%%
\subsection{Generating colored noise}
%%%%%%%%%%%%%%%%%%%%%%%%%%%%%%%%%%%%%%%%%%%%%%%%%%%%%%%%%%%%

\begin{figure}[!t]
\includegraphics[width=0.9\columnwidth]{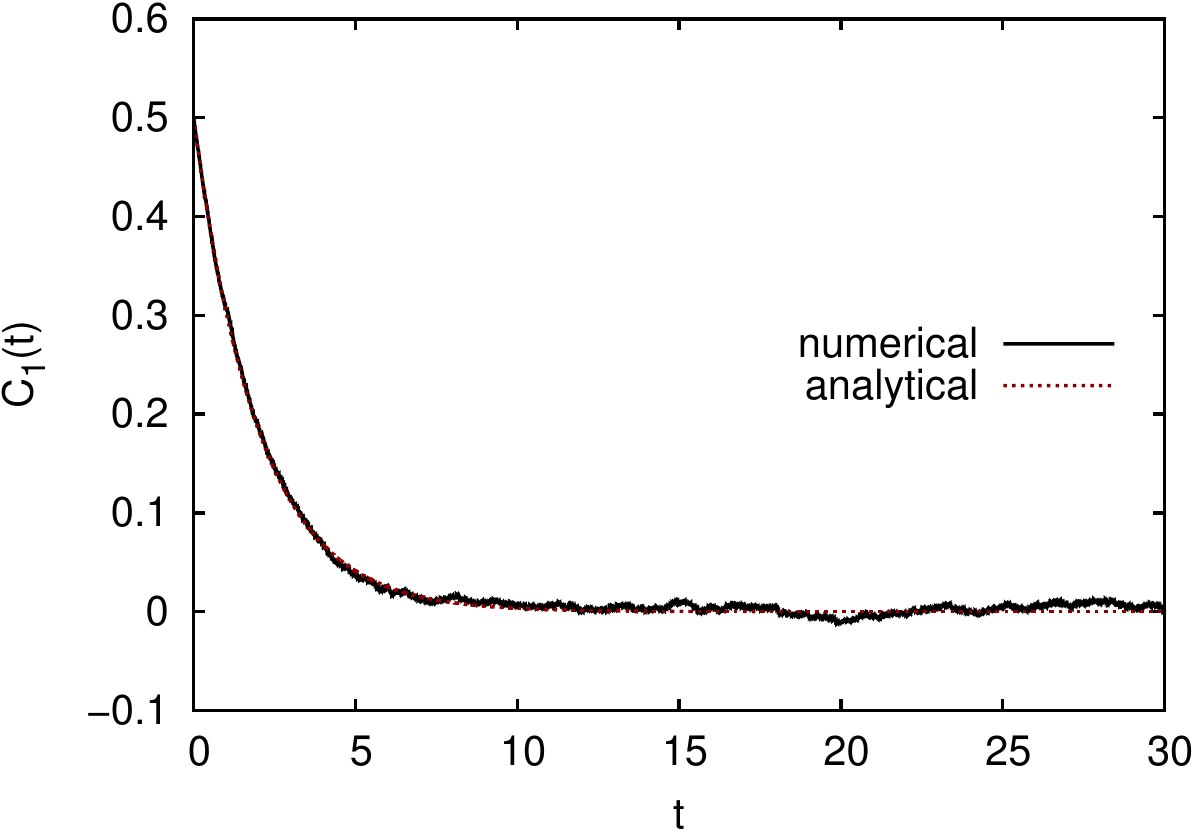}
\includegraphics[width=0.9\columnwidth]{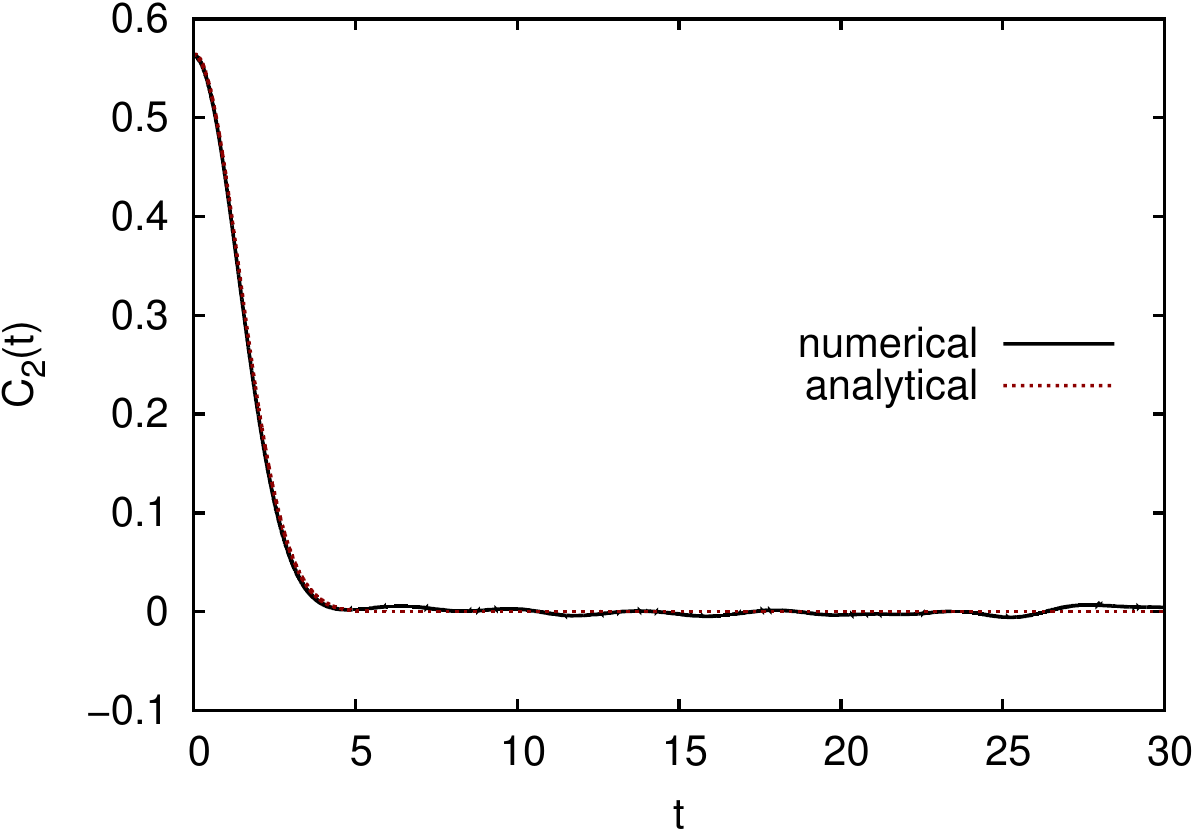}
\caption{(Color online) Comparison of the numerical simulation of
  $\langle \xi(0)\xi(t) \rangle$ with the analytical result for two
  different covariance functions ($C_1$, $D=2,\,\tau=2$ and $C_2$,
  $D=2,\,\alpha=2$) based on 8000 realizations.}
\label{fig:compareCovariance}
\end{figure}

We can modify our general equation of noise~\eqref{eq:generalNoise} in
such a way that it becomes related to the normalized white
noise~\eqref{eq:normWhiteNoise} via
\begin{equation}
\begin{split}
  \xi(t) &= \sum_{i=1}^n a_i b(t-t_i) \\
  &=  \sum_{i=1}^n \int_{-\infty}^{\infty} \! \mathrm{d}t' \, a_i b(t-t') \delta (t' - t_i) \\
  &= \int_{-\infty}^{\infty} \! \mathrm{d}t' \, b(t-t') \sum_{i=1}^n a_i \delta (t' - t_i) \\
  &= \int_{-\infty}^{\infty} \! \mathrm{d}t' \, b(t-t') \sigma\sqrt{\mu} \bar{\xi}_w(t') \\
  &= \int_{-\infty}^{\infty} \! \mathrm{d}t' \, G(t-t') \bar{\xi}_w(t')
  \,.
\end{split}
\end{equation}
With the substitution $t''= -t + t'$ and the symmetry of $G(t)$, caused
by its proportionality to the symmetric pulse shape $b(t)$, we obtain \cite{STOKASTISK}
\begin{equation}
\begin{split}
\label{eq:Noise}
\xi(t) &= \int_{-\infty}^{\infty} \! \mathrm{d}t'' \, G(t'') \bar{\xi}_w(t+t'') \\
		&= \int_{-\Delta}^{\Delta} \! \mathrm{d}t'' \, G(t'') \bar{\xi}_w(t+t'') \,.
\end{split}
\end{equation}
For practical reasons of generating stochastic variables at discrete
times with a constant time step $\Delta t$, a simple algorithm for the
discretized form of \eqref{eq:Noise} is presented in the appendix. We
verified this algorithm by comparing given covariance functions with the
numerical result of multiple realizations of this random process (cf.\
Fig.~\ref{fig:compareCovariance}). We used the following two covariance
functions and their Fourier transforms for the relations
\eqref{eq:covarianceNoise2} and \eqref{eq:covarianceNoise3}:
\begin{equation}
\begin{split}
\label{eq:covFunc}
  C_1(t) &= \frac{D}{2\tau} \exp \left (-\frac{|t|}{\tau} \right ) \;
  \\
  \Rightarrow\quad S_{\xi 1}(\omega)&=\frac{D}{1+\tau^2\omega^2}\,,\\
  C_{2}(t) &= \frac{D}{\alpha\sqrt{\pi}}
  \exp \left [- \left (\frac{t}{\alpha} \right)^2 \right ]\; \\
  \Rightarrow\quad S_{\xi 2}(\omega) &=D \exp \left [-\left (\frac{\alpha\omega}{2} \right)^2 \right]\,.
\end{split}
\end{equation}
Here, $\tau$ and $\alpha$ are positive values characterizing the
correlation time of the noise. For the limit $\tau \to 0$ and $\alpha
\to 0$, $C_1(t)$ and $C_2(t)$ approach the covariance function of the
white noise $D\delta(t)$.

According to (\ref{eq:covarianceNoise3}) we obtain:
\begin{equation}
\begin{split}
G_1(t) &=\frac{\sqrt{D}}{\pi \tau} \mathrm{K}_0 \left (\frac{|t|}{\tau}
\right ), \\
G_2(t) &= \frac{\sqrt{2D}}{\alpha \sqrt{\pi}} \exp \left
  (-\frac{2t^2}{\alpha^2} \right )\,,
\end{split}
\end{equation}
where $\mathrm{K}_0$ denotes the modified Bessel function of the second
kind.

In this way we have worked out a method to obtain a stationary Gaussian
process with an arbitrary covariance function and a positive-valued
Fourier transform. Using this approach, the noise can be simulated with
only small numerical effort (see the Appendix). The question is now how
the covariance function affects the solution of the Langevin equation.
%

%%%%%%%%%%%%%%%%%%%%%%%%%%%%%%%%%%%%%%%%%%%%%%%%%%%%%%%%%%%%%%%%%%%%%
\section{The generalized Langevin equation}
\label{sec:langevinColour}
%%%%%%%%%%%%%%%%%%%%%%%%%%%%%%%%%%%%%%%%%%%%%%%%%%%%%%%%%%%%%%%%%%%%

In the following we assume that the collisions experienced by an
observed particle in the heat bath are time-correlated with each other
resulting in a time-correlated noise for the Langevin equation. Since
the stochastic force as well as the friction force in the Langevin
equation are of the same origin, the friction force will also have a
time dependence. The environment of the particle is affected by its
movement and the particle is influenced to a later time in return, i.e.,
we describe a non-Markovian process with memory. Therefore, we introduce
a time-dependent friction kernel $\Gamma(t)$ leading to the following
form of the generalized one dimensional Langevin equation \cite{KUBO92}:
\begin{equation}
\label{eq:genLangevinEq}
m\ddot{x}(t)+2\int_0^t\!\mathrm{d}t'\,\Gamma(t-t')\dot{x}(t') - F_{\mathrm{ext}}(x) = \xi(t) \,,
\end{equation}
where $F_{\mathrm{ext}}(x)$ is an additional external force. Assuming
that the equipartition principle holds\footnote{Here and in the
  following we set the Boltzmann constant $\kB=1$.},
\begin{equation*}
\frac{1}{2}m \langle v^2 \rangle = \frac{1}{2} T \,,
\end{equation*}
we obtain a relation between the friction kernel $\Gamma(t)$ and the
covariance function $C(t)$:
\begin{equation}
\label{eq:flucDiss}
\Gamma(t-t')=\frac{1}{2 T} \langle \xi(t)\xi(t') \rangle =\frac{1}{2 T}C(t-t') \,.
\end{equation}
This is the well known fluctuation-dissipation theorem. For an arbitrary
time-independent external force $F_{\mathrm{ext}}(x)$ a derivation can
be found in \cite{CORTES}.

Our numerical solving algorithm for the generalized Langevin equation
\eqref{eq:genLangevinEq} is based on the three-step Adams-Bashforth
scheme \cite[p.~307]{NUMERICAL}, where the right side is calculated with
the method presented in the Appendix.  Therefore, for each particle we
store the position $x(t)$ and velocity $\dot{x}(t)$ at every time step.

%%%%%%%%%%%%%%%%%%%%%%%%%%%%%%%%%%%%%%%%%%%%%%%%%%%%%%%%%%%%%%%%%%%%%%%%%%
\section{Langevin equation without external potential}
\label{sec:noPotential}
%%%%%%%%%%%%%%%%%%%%%%%%%%%%%%%%%%%%%%%%%%%%%%%%%%%%%%%%%%%%%%%%%%%%%%%%%%

We first analyze the generalized Langevin equation without an external
potential,
\begin{equation}
\label{eq:colourLangevinNoPot}
m\ddot{x}(t)+2\int_0^t\!\mathrm{d}t'\,\Gamma(t-t')\dot{x}(t') = \xi(t) \,,
\end{equation}
\begin{figure}[htp]
  \includegraphics[width=0.9\columnwidth]{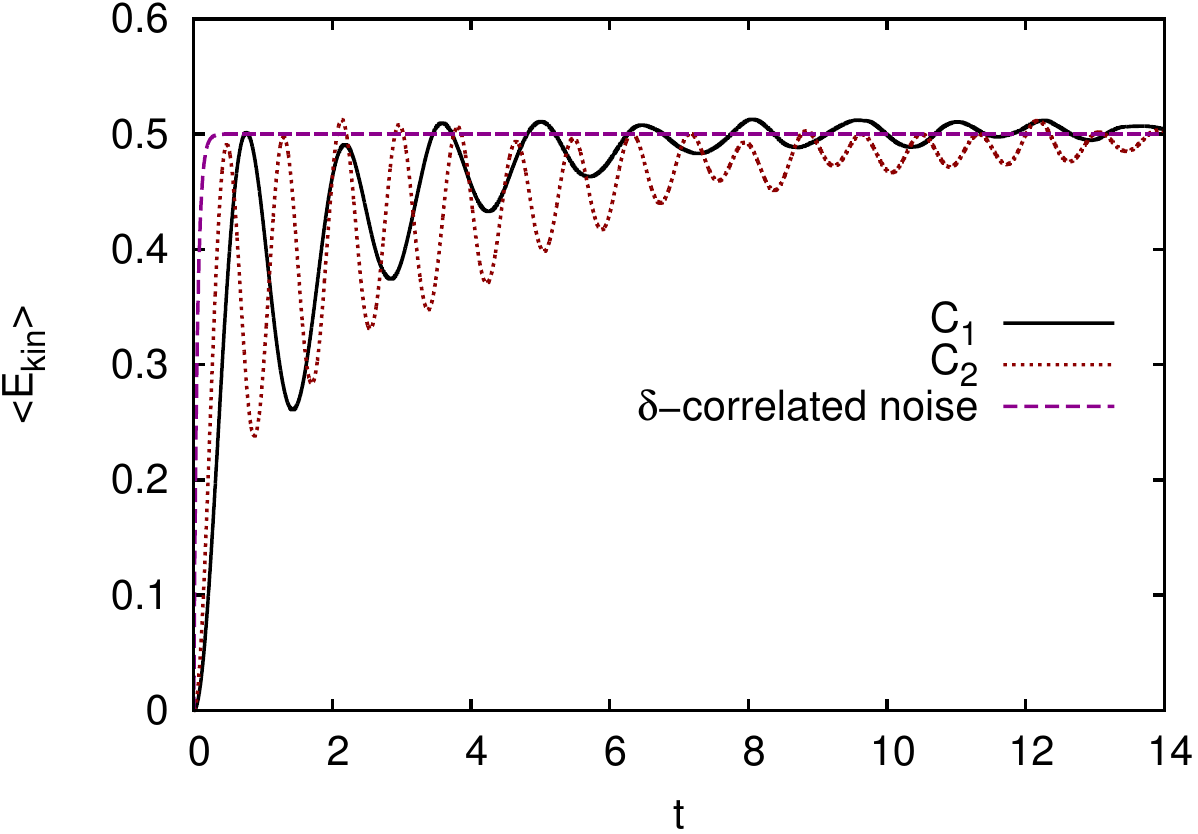}
  \caption{(Color online) Numerical evolution of the mean kinetic energy
    \eqref{eq:colourLangevinNoPot} for the exponential covariance
    function $C_1(t)$ with $\tau=2$, the Gaussian covariance function
    $C_2(t)$ with $\alpha=1$, and the theoretical limit of $\tau\to 0$,
    $\alpha\to 0$ corresponding to $\delta$-correlated white noise. The
    initial conditions for both simulations are given
    in~\eqref{eq:pot0Init2}.}
  \label{fig:pot0Cor0Cor1}
\end{figure}
where for initial conditions we choose
\begin{equation}
\label{eq:pot0Init2}
x_0 = 0, \quad v_0 = 0, \quad D=2, \quad m=0.1, \quad T = 1\,.
\end{equation}
We note that these initial conditions imply that this ``Brownian
particle'' is not assumed to be in equilibrium with the ``heat bath''
represented by the fluctuating force, $\xi$, which is described as
stationary Gaussian noise. Consequently the solutions of
(\ref{eq:colourLangevinNoPot}) includes transient motion of the Brownian
particle, while the corresponding back reaction to the heat bath is
neglected. Indeed, for both covariance functions, $C_1(t)$ and $C_2(t)$,
we obtain an oscillating transient solution of the Langevin equation
until the equilibrium value is reached, which differs significantly from
the exponential trend in case of $\delta$-correlated white noise (see
Fig.~\ref{fig:pot0Cor0Cor1})\cite{Kubo:1966fluctdis}. The oscillation is
due to the retarded friction on the particle due to the memory of the
system.

%%%%%%%%%%%%%%%%%%%%%%%%%%%%%%%%%%%%%%%%%%%%%%%%%%%%%%%%%%%%
\subsection{Analytical solution for \texorpdfstring{$C_1$}{C1}}
%%%%%%%%%%%%%%%%%%%%%%%%%%%%%%%%%%%%%%%%%%%%%%%%%%%%%%%%%%%%

To solve the generalized Langevin
equation~\eqref{eq:colourLangevinNoPot} analytically as a linear
integro-differential equation, we can benefit from a continuation of the
velocity $v(t)$ by defining $v_T$ as
\begin{equation}
v_T(t)=
	\begin{cases}
          0 \quad &\mathrm{for}\quad t<0\,, \\
          v(t) \quad &\mathrm{for}\quad t \in [0,T]\,,\\
          v(T) \quad &\mathrm{for}\quad t > T
	\end{cases}
\end{equation}
and choosing $v(0) = 0$, such that $v_T(t)$ is a continuous function.
With
\begin{equation*}
\xi_T(t)=
	\begin{cases}
          0 \quad &\mathrm{for} \quad t<0\,, \\
          \xi(t) \quad &\mathrm{for} \quad t \in [0,T]\,,\\
          0 \quad &\mathrm{for} \quad t > T\\
	\end{cases}
\end{equation*}
and
\begin{equation*}
\Gamma_{\text{ret}}(t)=
	\begin{cases}
		0 \quad &\mathrm{for} \quad t<0\,, \\
		\Gamma(t) \quad &\mathrm{for} \quad t > 0\\
	\end{cases}
\end{equation*}
we obtain an equation which is identical to
Eq.~\eqref{eq:colourLangevinNoPot} in the interval $t\in[0,T]$:
\begin{equation*}
m\dot{v}_T(t)+2\int_{-\infty}^{\infty}\!\mathrm{d}t'\,\Gamma_{\mathrm{ret}}(t-t')v_T(t')=\xi_T(t) \,.
\end{equation*}
This linear differential equation can be solved with the ansatz
\begin{equation}
\label{eq:greenDef}
m\dot{G}_{\mathrm{ret}}(t)+2\int_{-\infty}^{\infty}\!\mathrm{d}t'\,\Gamma_{\mathrm{ret}}(t-t')G_{\mathrm{ret}}(t')=\delta(t) \,,
\end{equation}
where $G_{\mathrm{ret}}(t)$ denotes a retarded Green's function. The
velocity is then given by
\begin{equation}
\label{eq:greenSolution}
v_T(t) = \int_{-\infty}^{\infty} \! G_{\mathrm{ret}}(t-t')\xi_T(t')
\mathrm{d}t' \quad \text{for} \quad t\in[0,T]\,.
\end{equation}
The Green's function can be found via a Fourier transform of
(\ref{eq:greenDef}), leading to
\begin{equation}
\label{eq:greenDefFour}
\tilde {G}_{\text{ret}}(\omega) = \frac{1}{2
  \tilde{\Gamma}_{\text{ret}}(\omega)-\ii m
  \omega}.
\end{equation}
For the exponential covariance function, $C_1(t)$, we find
\begin{equation}
\label{green.3}
\tilde{\Gamma}_{\text{ret}}(\omega)=\frac{D}{4 T} \frac{1}{1-\ii
  \omega \tau}
\end{equation}
and thus for the retarded Green's function, according to
(\ref{eq:greenDefFour})
\begin{equation}
\begin{split}
\label{green.4}
   \tilde{G}_{\text{ret}}(\omega)
  &=\frac{1}{m \tau} \frac{\ii \omega \tau -1}{\omega^2+\ii \omega / \tau
    - Q/(2 \tau)},
\end{split}
\end{equation}
where $Q = \frac{D}{m T}$.  The Fourier transformation to the time
domain is done in the usual way, using the theorem of residues by
closing the integration path in the upper (lower) $\omega$-half plane
for $t<0$ ($t>0$). As to be expected from the retardation condition,
$\tilde{G}_{\text{ret}}$ is analytic in the upper half-plane. For $t>0$
and $2 Q \tau<1$, defining $\gamma_c=\frac{\sqrt{1-2 Q \tau}}{2 \tau}$,
the Green's function reads
\begin{equation}
\label{green.4.2}
  G_{\mathrm{ret}}(t) = \frac{1}{m}\left [ \frac{1}{2\gamma_c\tau} \sinh(\gamma_c t)
    + \cosh(\gamma_c t) \right ] \ee^{-\frac{t}{2\tau}}.
\end{equation}
This can be analytically continued to the case $2Q\tau > 1$ by setting
$\gamma_c=\ii \omega_c$ with $\omega_c=\frac{\sqrt{2 Q \tau-1}}{2 \tau}$:
\begin{equation}
  G_{\mathrm{ret}}(t) =  \frac{1}{m}\left [ \frac{1}{2\omega_c\tau} \sin(\omega_c t)
    + \cos(\omega_c t) \right ] \ee^{-\frac{t}{2\tau}}.
\end{equation}
Finally, the case $2 Q \tau=1$ can be found by taking the limit
$\gamma_c \rightarrow 0$ of (\ref{green.4.2}), resulting in
\begin{equation}
G_{\text{ret}}(t)=\frac{1}{m} \left (\frac{t}{2 \tau}+1 \right )
\ee^{-\frac{t}{2 \tau}}.
\end{equation}
For small correlation times, i.e., for $2 Q \tau \leq 1$ we find an
exponential-decay behavior. The relaxation time is larger compared to
the Markovian limit due to the memory effects described by the
correlation function. For larger correlation times the system oscillates
with the characteristic frequency $\omega_c =
\frac{\sqrt{2Q\tau-1}}{2\tau}$ due to the memory of the medium, leading
to a kind of ``plasmon formation''. This is also reflected in the
velocity-correlation function, which we evaluate next.

Inserting the Fourier transform of $\tilde{v}_T(\omega)$ from
Eq.~\eqref{eq:greenSolution},
\begin{equation}
\tilde{v}_T(\omega)=\tilde{G}_{\text{ret}}(\omega) \tilde{\xi}_T(\omega)
\end{equation}
into the definition of the velocity's spectral density \cite[p.~60]{TLE}
yields
\begin{equation}
\begin{split}
  S_v(\omega) &= \lim_{T \to \infty} \frac{1}{T}\left\langle {\lvert\tilde{v}_T(\omega)\rvert}^2 \right\rangle \\
  &= \lim_{T \to \infty} \frac{1}{T} {\lvert\tilde{G}_{\mathrm{ret}}(\omega)\rvert} ^2 \left\langle {\lvert\tilde{\xi}_T(\omega)\rvert}^2 \right\rangle \\
  &= {\lvert\tilde{G}_{\mathrm{ret}}(\omega)\rvert} ^2 S_{\xi}(\omega)=2
  T {\lvert\tilde{G}_{\mathrm{ret}}(\omega)\rvert} ^2 \tilde{\Gamma}(\omega)
  \,.
\end{split}
\end{equation}

\begin{figure}[htp]
\vspace*{2mm}
\includegraphics[width=0.95 \columnwidth]{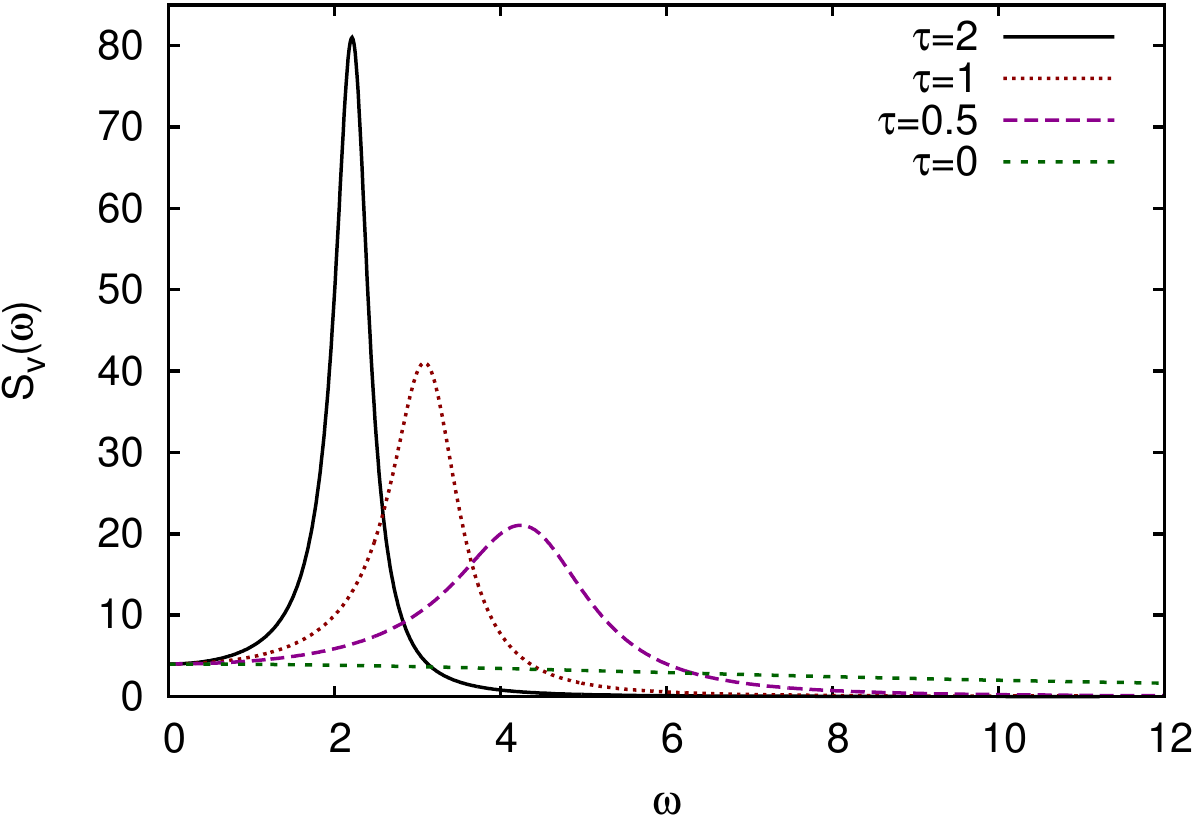}
\caption{(Color online) The spectral density of the velocity for
  different correlation times. For small values of $\tau$ the peak
  becomes broader and is shifted to higher frequencies.}
\label{fig:pot0SpectralDensity}
\end{figure}

Since $\Gamma(t)=\Gamma(-t) \in \R$, using the definition of the
retarded damping function a direct evaluation of its Fourier transform
yields ($\omega \in \R$)
\begin{equation}
\tilde{\Gamma}(\omega)=\tilde{\Gamma}_{\text{ret}}(\omega)+\tilde{\Gamma}_{\text{ret}}^*(\omega).
\end{equation}
Using (\ref{green.4}) this can be written as
\begin{equation}
\tilde{\Gamma}=\frac{1}{2} \left
  (\frac{1}{\tilde{G}_{\text{ret}}}+\frac{1}{\tilde{G}_{\text{ret}}^*}
\right)=\frac{\re \tilde{G}_{\text{ret}}}{|\tilde{G}_{\text{ret}}|^2},
\end{equation}
and with the fluctuation-dissipation relation (\ref{eq:flucDiss}) we
finally arrive at
\begin{equation}
\begin{split}
S_v(\omega) =& 2 T \, \re \left[ \tilde{G}_{\mathrm{ret}}(\omega) \right] \\
	=& \frac{D^2}{m^2}\,\frac{1}{\frac{1}{4}Q^2+(1-Q\tau)\omega^2+\tau^2\omega^4}.
\end{split}
\end{equation}

Fig.~\ref{fig:pot0SpectralDensity} shows the spectral density for
different correlation times $\tau$, where for long correlation times we
observe a clear oscillation expressed by a sharp peak at the frequency
$\omega_{\mathrm{peak}}$, which is given by
\begin{equation*}
\omega_{\mathrm{peak}} = \frac{\sqrt{2Q\tau-2}}{2\tau} \,.
\end{equation*}
If $\tau$ is large enough, it follows that $\omega_{\mathrm{peak}}
\approx \frac{\sqrt{2Q\tau-1}}{2\tau} = \omega_c$. For decreasing $\tau$
the peak becomes broader, and its maximum moves to the right until
$\omega_{\mathrm{peak}}$ reaches an extremum for $Q\tau=2$. For very
small values of $\tau$ the maximum of the spectral density remains at
$0$ and approaches a Lorentz shape with a width of $\frac{\gamma}{m}$
\cite[p.~53]{XU1}, \cite{Xu:1999aq}, where $\gamma=\frac{D}{2 T}$
according to the Nernst-Einstein relation. As expected, this leads to
the Markovian limit for the Langevin equation with white noise.

For $2Q\tau > 1$ we obtain an expression for the mean kinetic energy
$\left\langle E_{\mathrm{kin}}(t) \right\rangle$:
\begin{widetext}
\begin{equation}
\begin{split}
\label{eq:pot0KinEnergy}
\erw{E_{\text{kin}}}&=\frac{1}{2}m \left\langle v^2(t) \right\rangle =
\frac{1}{2}m \int_0^t \!\mathrm{d}s \int_0^t \!\mathrm{d}s' \,
G_{\mathrm{ret}}(t-s)G_{\mathrm{ret}}(t-s') \left\langle \xi(s)\xi(s')
\right\rangle \\
&= \frac{1}{2}m \int_0^t \!\mathrm{d}s \int_0^t \!\mathrm{d}s'
G_{\mathrm{ret}}(t-s)G_{\mathrm{ret}}(t-s') C(s-s') \\
&= \frac{1}{2} T - \frac{1}{2}\frac{T}{2 Q \tau - 1} \left[
  Q\tau + \sqrt{2Q\tau-1} \sin(2\omega_c t) + (Q\tau - 1) \cos(2\omega_c
  t) \right] \ee^{-\frac{t}{\tau}},
\end{split}
\end{equation}
\end{widetext}
showing the relaxation to the equilibrium value $T/2$ with the
damping time $\tau$ and oscillations due to the memory effect.

%%%%%%%%%%%%%%%%%%%%%%%%%%%%%%%%%%%%%%%%%%%%%%%%%%%%%%%%%%%%
\subsection{Numerical results}
%%%%%%%%%%%%%%%%%%%%%%%%%%%%%%%%%%%%%%%%%%%%%%%%%%%%%%%%%%%%

\begin{figure}[htp]
\includegraphics[width=0.9\columnwidth]{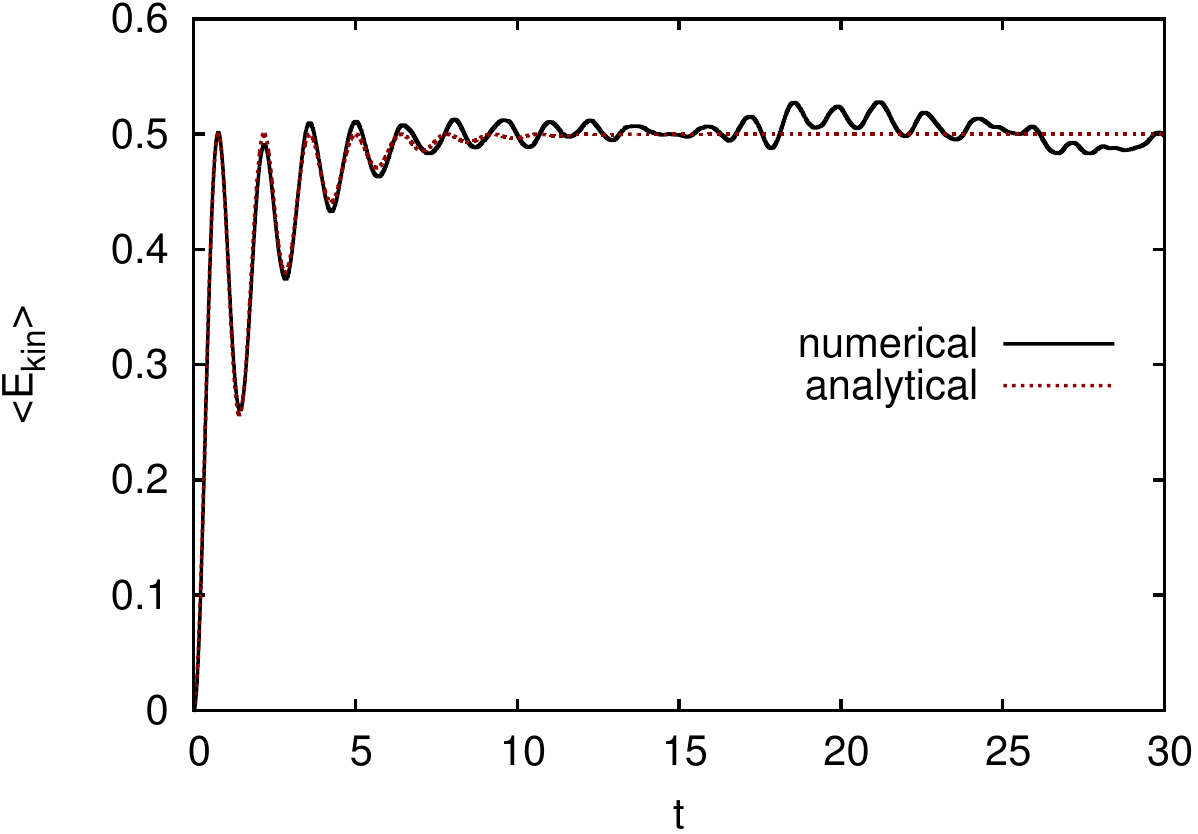}
\caption{(Color online) Comparison of the kinetic energy between the
  analytical expression (Eq.~\eqref{eq:pot0KinEnergy}) and the numerical
  average over 8000 realizations.}
\label{fig:pot0KinEnergyTheoEx}
\end{figure}

\begin{figure}[htp]
\includegraphics[width=0.9\columnwidth]{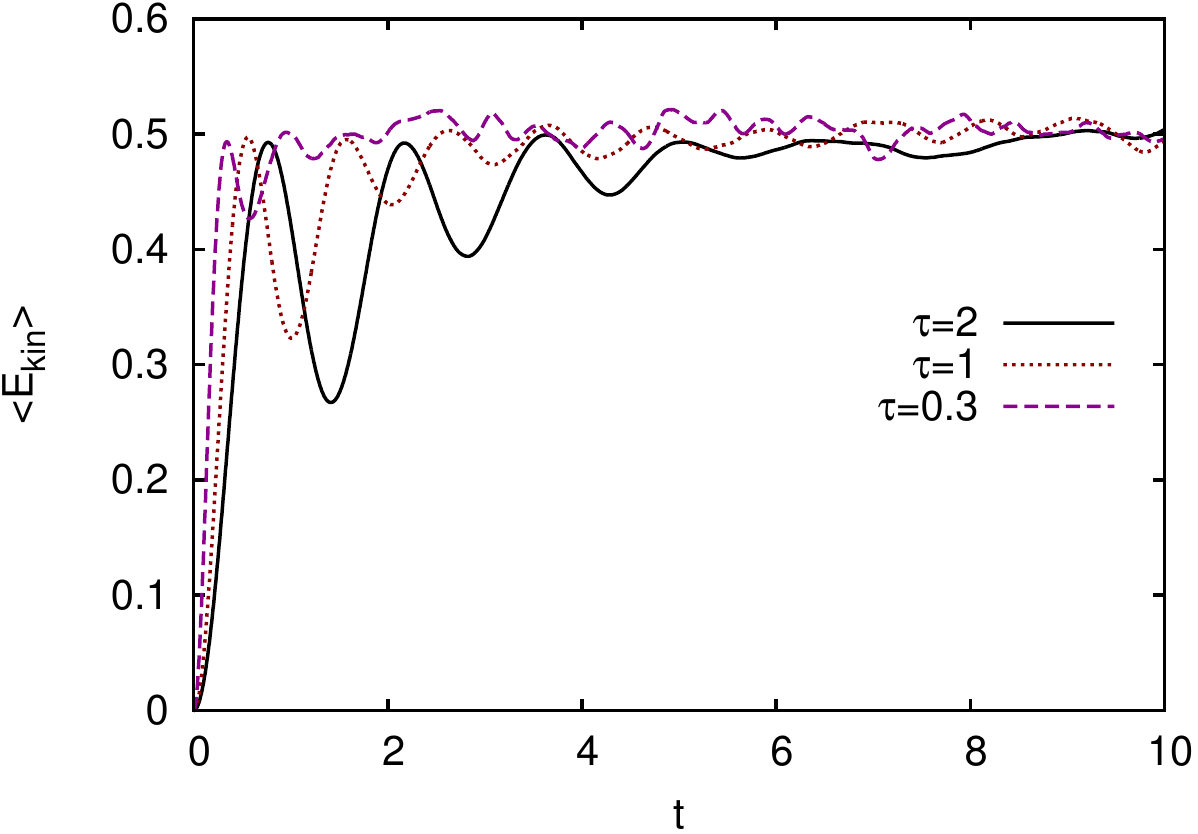}
\caption{(Color online) Numerical results for the kinetic energy and different values
  of $\tau$. As expected, with increasing correlation times the system
  takes longer to approach the equilibrium state.}
\label{fig:pot0KinEnergy}
\end{figure}

In Fig.~\ref{fig:pot0KinEnergyTheoEx} we compare the analytical
expression of the kinetic energy with the numerical average over 8000
realizations, where the initial conditions are given by
\begin{equation*}
x_0 = 0, \quad v_0 = 0, \quad D=2, \quad m=0.1, \quad T = 1
\end{equation*}
and $\tau=2$. Our numerical simulation is in a good agreement with the
analytical result. Fig.~\ref{fig:pot0KinEnergy} shows the kinetic energy
for different values of $\tau$. For a system in equilibrium, we have
verified numerically that the velocity of particles is Boltzmann
distributed.


%%%%%%%%%%%%%%%%%%%%%%%%%%%%%%%%%%%%%%%%%%%%%%%%%%%%%%%%%%%%%%%%%%
\section{Langevin equation in quadratic potentials}
\label{sec:harmonic}
%%%%%%%%%%%%%%%%%%%%%%%%%%%%%%%%%%%%%%%%%%%%%%%%%%%%%%%%%%%%%%%%%%

In this Section we verify our numerical algorithm to simulate
non-Markovian Brownian motion for the analytically solvable case of the
motion in quadratic potentials, i.e., the harmonic-oscillator and the
quadratic-barrier potential.

\subsection{Harmonic-oscillator potential}

\begin{figure}[htp]
  \includegraphics[width=0.9\columnwidth]{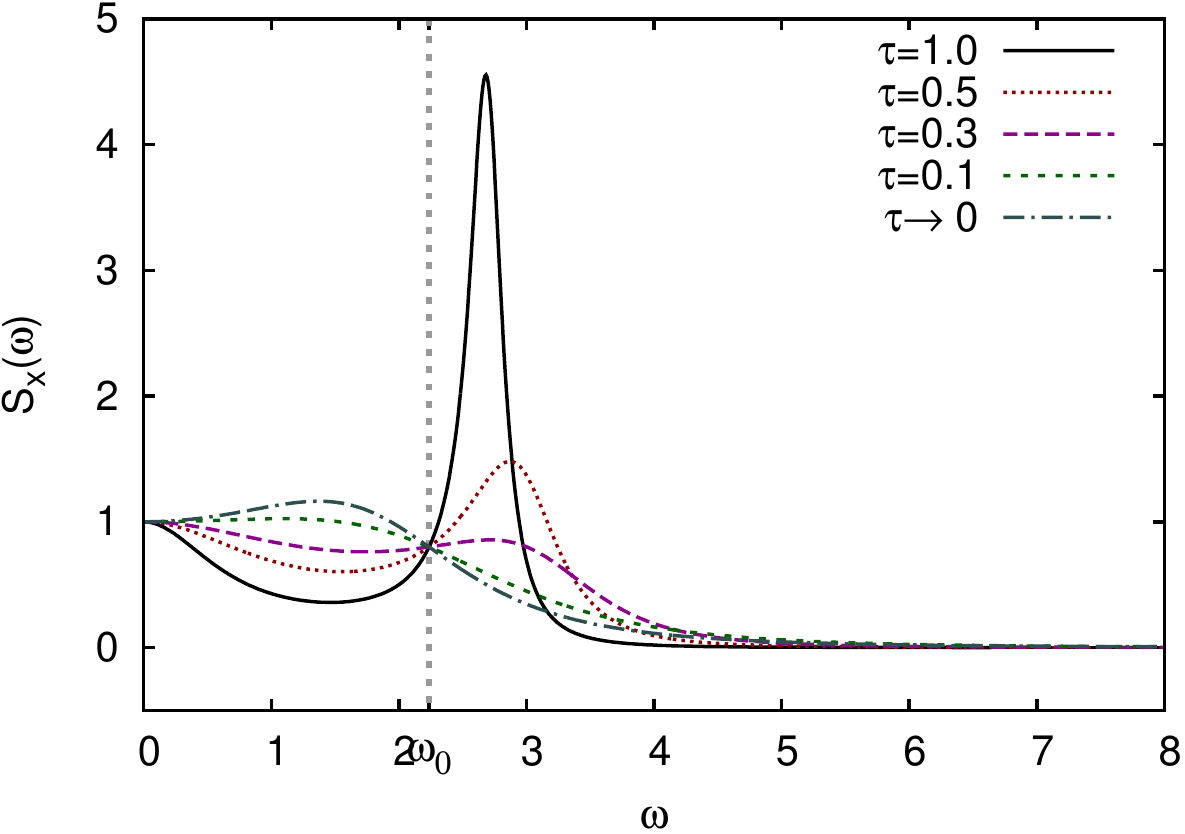}
  \caption{(Color online) Spectral density of the position $x$ in a harmonic
    oscillator. The simulation parameters are $T = 1$, $D=1$,
    $m=0.2$ and $\omega_0 = \sqrt{5}$.}
  \label{fig:pot1Sx}
\end{figure}

\begin{figure}[htp]
  \includegraphics[width=0.9\columnwidth]{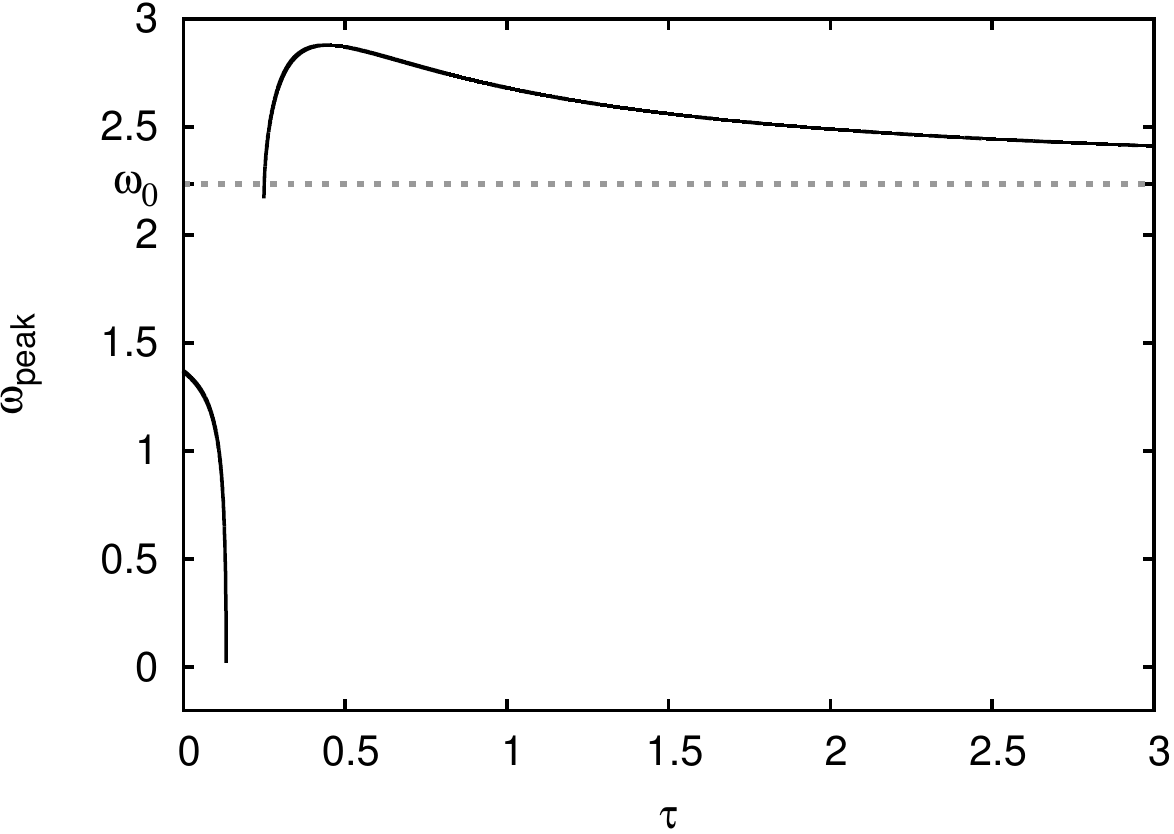}
  \caption{(Color online) Behavior of the peak frequency
    $\omega_\mathrm{peak}$ as a function of $\tau$. In a narrow
    frequency range the peak vanishes completely. With increasing values
    of $\tau$ it approaches the frequency of the harmonic oscillator
    whereas in the limit $\tau \to 0$ its frequency converges to the
    white-noise limit. The same simulation parameters have been used as
    in Fig.~\ref{fig:pot1Sx}.}
\label{fig:pot1wPeak}
\end{figure}

We now add a harmonic potential
\begin{equation}
V_{\mathrm{ext}}(x)=\frac{1}{2}m\omega_0^2x^2
\end{equation}
as a minimal extension to include an external force. In this case the
Langevin equation takes the form:
\begin{equation}
m\ddot{x}(t) + 2\int_0^t\!\mathrm{d}t'\,\Gamma(t-t')\dot{x}(t') +
m\omega_0^2 x = \xi(t) \,.
\end{equation}
From the equipartition and virial theorems we expect
\begin{equation}
\frac{1}{2}m \left\langle v^2(t) \right\rangle = \frac{1}{2} m
\omega_0^2 \left\langle x^2(t) \right\rangle = \frac{1}{2} T \,.
\end{equation}
For the initial conditions $x_0=0$ and $v_0=0$ we can calculate the
spectral density of the position $x$ in analogy to the spectral density
of the velocity without potential,
\begin{equation*}
S_x(\omega)={\lvert\tilde{G}_{\mathrm{ret}}(\omega)\rvert}^2 S_{\xi}(\omega) \,.
\end{equation*}
Here, $\tilde{G}_{\mathrm{ret}}(\omega)$ denotes the retarded Green's
function of $x_T(t)$ which solves the equation
\begin{equation}
\begin{split}
m\ddot{G}_{\mathrm{ret}}(t) + 2\int_{-\infty}^{\infty}\!\mathrm{d}t'\,\Gamma_{\mathrm{ret}}(t-t')\dot{G}_{\mathrm{ret}}(t')& \\
+ m\omega_0^2 G_{\mathrm{ret}}(t) = \delta(t)& \,.
\end{split}
\end{equation}
It can be evaluated in an analogous way as for the free particle cf.\
Sec.\ \ref{sec:noPotential}, leading to
\begin{equation}
S_x(\omega) =\frac{2T}{\omega} \mathrm{Im}[\tilde{G}_{\mathrm{ret}}(\omega)]
\end{equation}
with
\begin{equation}
\tilde{G}_{\mathrm{ret}}(\omega) =
\frac{1}{m\left(\omega_0^2-\omega^2-i\frac{2\omega}{m}\tilde{\Gamma}_{\mathrm{ret}}(\omega)
  \right)}\,.
\end{equation}
Fig.~\ref{fig:pot1Sx} shows a peak in the spectral density approaching
the frequency $\omega_0$ for long correlation times $\tau$. For smaller
values of $\tau$ the peak becomes broader. The frequency of this peak is
denoted with $\omega_\mathrm{peak}$, and its development as a function
of $\tau$ is shown in Fig.~\ref{fig:pot1wPeak}. In the present example
the damping of the system, characterized by $D$, is relatively
large. With decreasing values of $D$ the function of the peak frequency
becomes continuous.

\subsection{Diffusion over a barrier}

\begin{figure}[htp]
  \includegraphics[width=0.9\columnwidth]{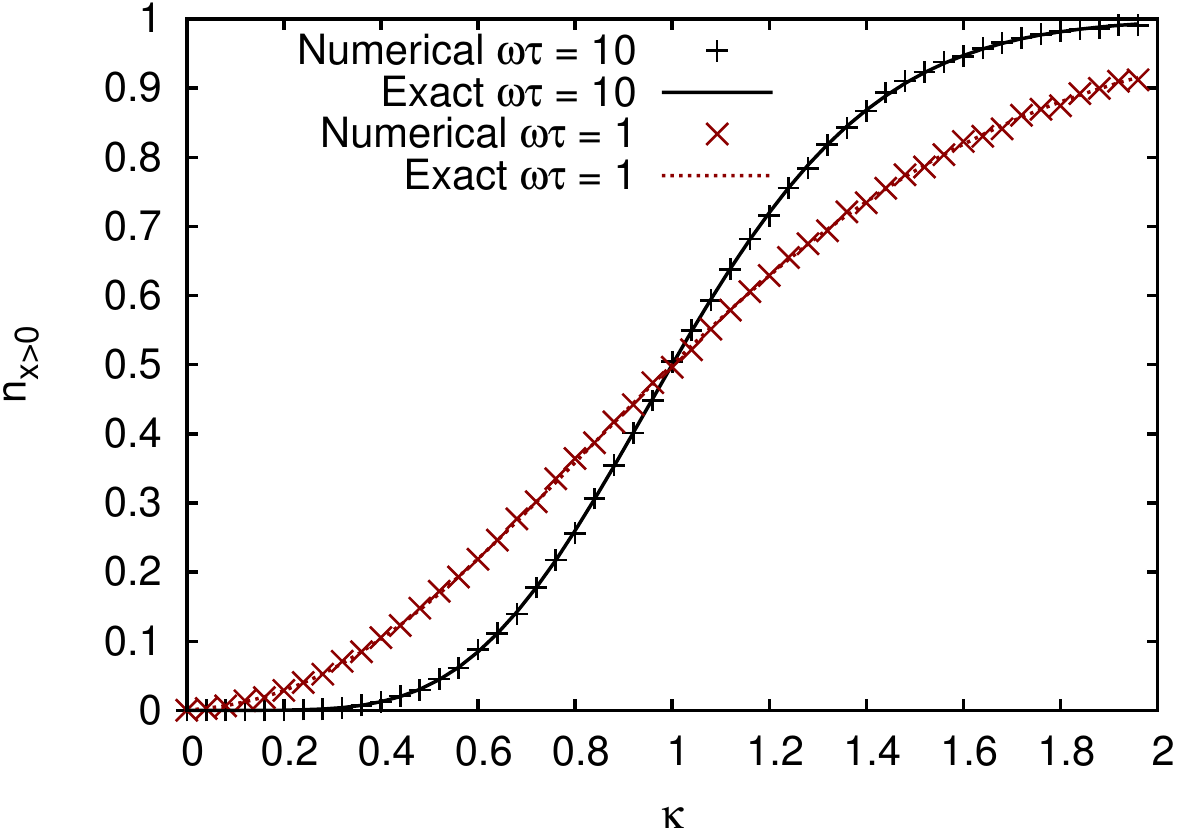}
  \caption{(Color online) Simulation for the probability to pass a
    square barrier compared to the analytic result (\ref{barr.2})
    \cite{Boilley:2006mw}. We use the same representation of the barrier
    height as in this reference, i.e., $\kappa=K/B_{\text{eff}}$ with
    $B_{\text{eff}}=B \omega_0/\lambda_1^2$.}
\label{fig:barrier}
\end{figure}
As an additional test of our numerical method, we simulate the diffusion
of a non-Markovian Brownian particle over a square-barrier potential,
\begin{equation}
\label{barr.1}
V(x)=-\frac{m \omega_0^2}{2} x^2.
\end{equation}
As detailed in \cite{Boilley:2006mw}, this problem can be solved
analytically for the case of the correlation function $C_1$
(\ref{eq:covFunc}). Although the Brownian particle can not come to
thermal equilibrium in this case, because the potential is not bounded
from below, the probability to pass over the barrier is well
defined. With the initial kinetic energy $K=p_0^2/2m$ ($p_0>0$) and the
barrier height $B=m \omega_0^2 x_0^2/2$ ($x_0<0$), $\beta=D/(2mT)$ in the
here simulated case of a stationary non-Markovian Langevin process it
reads for $t \rightarrow \infty$
\begin{equation}
\begin{split}
\label{barr.2}
P(x_0,p_0)=\frac{1}{2} \mathrm{erfc} \Bigg
  [ \omega_0 &\sqrt{\frac{1+\lambda_1 \tau}{\beta \lambda_1}}  \\
&\times \Bigg (\sqrt{\frac{B}{T}}
- \frac{\lambda_1}{\omega_0} \sqrt{\frac{K}{T}}
      \Bigg ) \Bigg ].
\end{split}
\end{equation}
The parameter $\lambda_1$ is the positive root of the cubic equation
\begin{equation}
\label{barr.3}
\lambda^3 + \frac{\lambda^2}{\tau}+\left(\frac{\beta}{\tau}-\omega_0^2
\right) -\frac{\omega_0^2}{\tau}=0.
\end{equation}
To further validate our numerics, we have used the same test cases as in
\cite{Boilley:2006mw}. As can be seen in Fig.\ \ref{fig:barrier}, the
results of the simulation is in perfect agreement with the analytical
result (\ref{barr.2}). We have checked that further evolution to later
times within our numerical simulation does not change the passing
probability anymore, i.e., that the time evolution really converges to
the anlytical result. We have also verified that the same results can be
achieved with the stochastic process, using a white-noise auxiliary
variable, as explained in the reference.

%%%%%%%%%%%%%%%%%%%%%%%%%%%%%%%%%%%%%%%%%%%%%%%%%%%%%%%%%%%%
\section{Langevin equation in a double well potential}
\label{sec:langevinDoubleWell}
%%%%%%%%%%%%%%%%%%%%%%%%%%%%%%%%%%%%%%%%%%%%%%%%%%%%%%%%%%%%
\begin{figure}[htp]
\includegraphics[width=0.9\columnwidth]{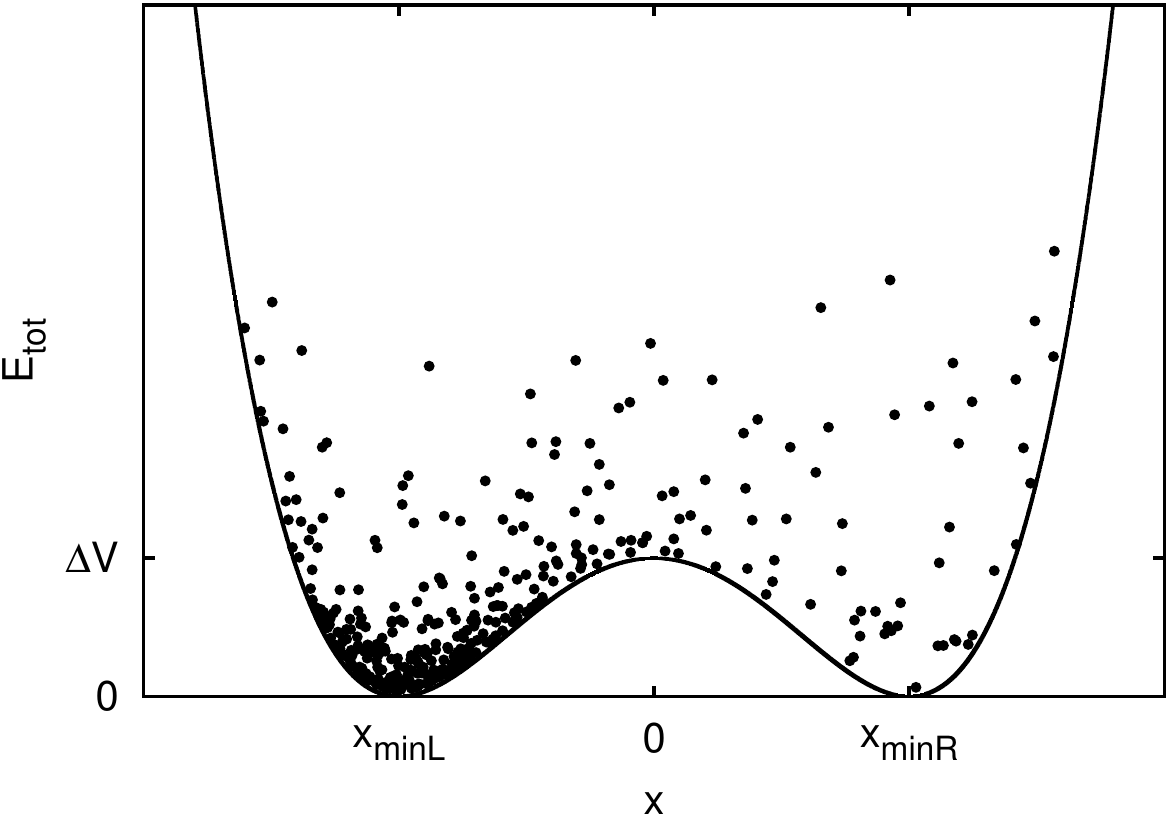}
\caption{(Color online) 400 independent realizations of the Langevin
  equation~\eqref{eq:langevinDoubleWell} with a symmetric double-well
  potential and initial conditions as defined
  in~\eqref{eq:pot2Init}. Shown is the total energy and position of each
  particle at the time $t=5$.}
\label{fig:chapterStarterPot2}
\end{figure}
The general form of a double-well potential $V(x)$ is described by
\begin{equation*}
V(x)=ax^4+bx^3+cx^2+d
\end{equation*}
with a suitable choice of parameters $a$, $b$, $c$ and $d$. Since the
corresponding force is not linear in $x$, a linear Green's function
method is no longer applicable to solve the Langevin equation
\begin{equation}
\begin{split}
\label{eq:langevinDoubleWell}
m\ddot{x}(t) &+ 2\int_0^t\!\mathrm{d}t'\,\Gamma(t-t')\dot{x}(t') \\
&+ 4ax^3 + 3bx^2 + 2cx = \xi(t) \,,
\end{split}
\end{equation}
and we present only numerical results.

%%%%%%%%%%%%%%%%%%%%%%%%%%%%%%%%%%%%%%%%%%%%%%%%%%%%%%%%%%%%
\subsection{Symmetric double well potential}
\label{sec:langevinDoubleWellSym}
%%%%%%%%%%%%%%%%%%%%%%%%%%%%%%%%%%%%%%%%%%%%%%%%%%%%%%%%%%%%

\begin{figure}[htp]
  \includegraphics[width=0.9\columnwidth]{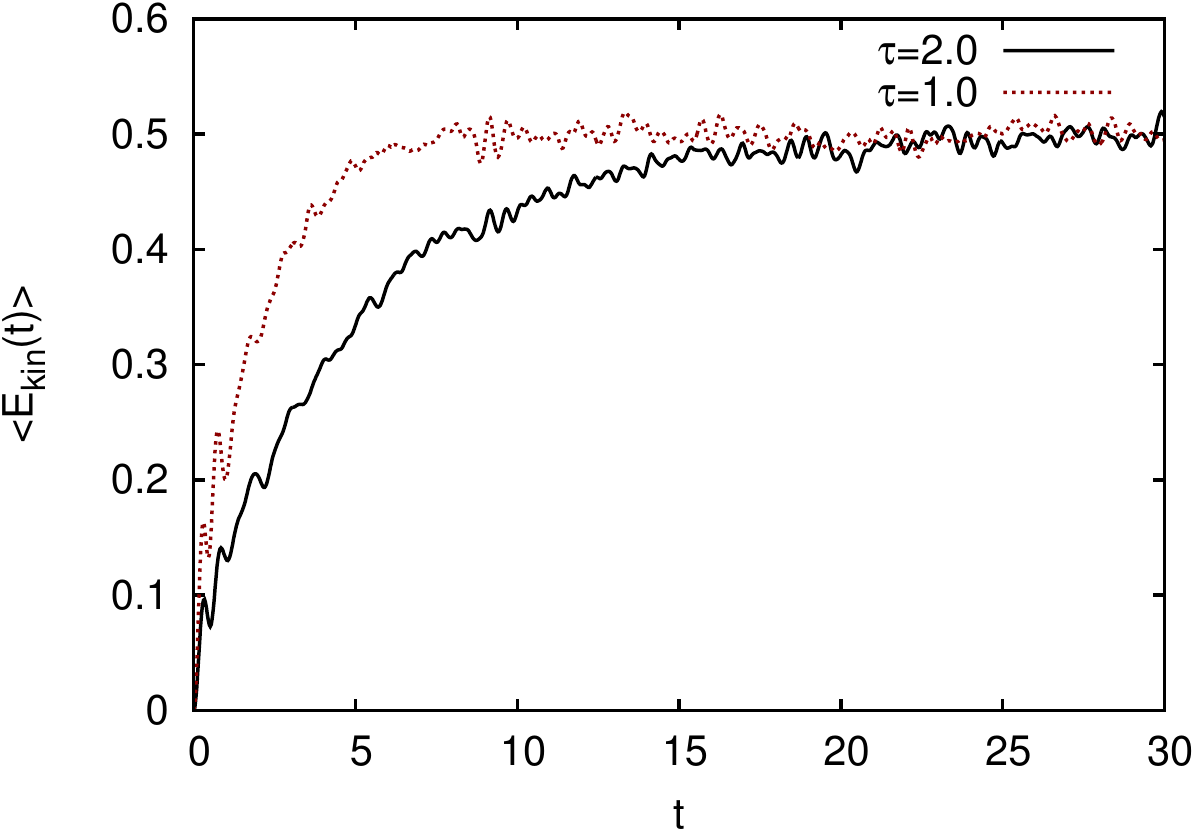}
  \caption{(Color online) Kinetic energy of the Langevin
    equation~\eqref{eq:langevinDoubleWell} with a symmetric double well
    potential for the covariance function $C_1$.}
  \label{fig:pot2kinEnergyTau}
	\end{figure}

\begin{figure}[htp]
\includegraphics[width=0.9\columnwidth]{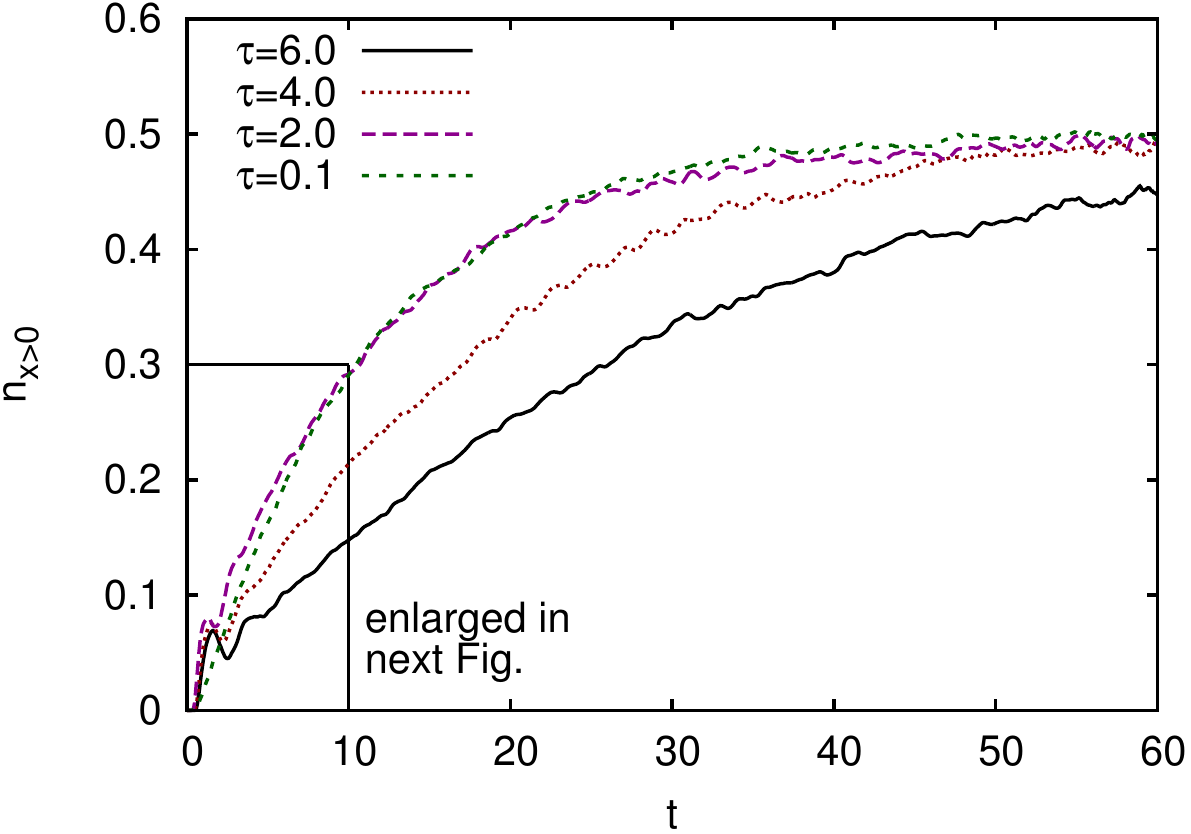}
\includegraphics[width=0.9\columnwidth]{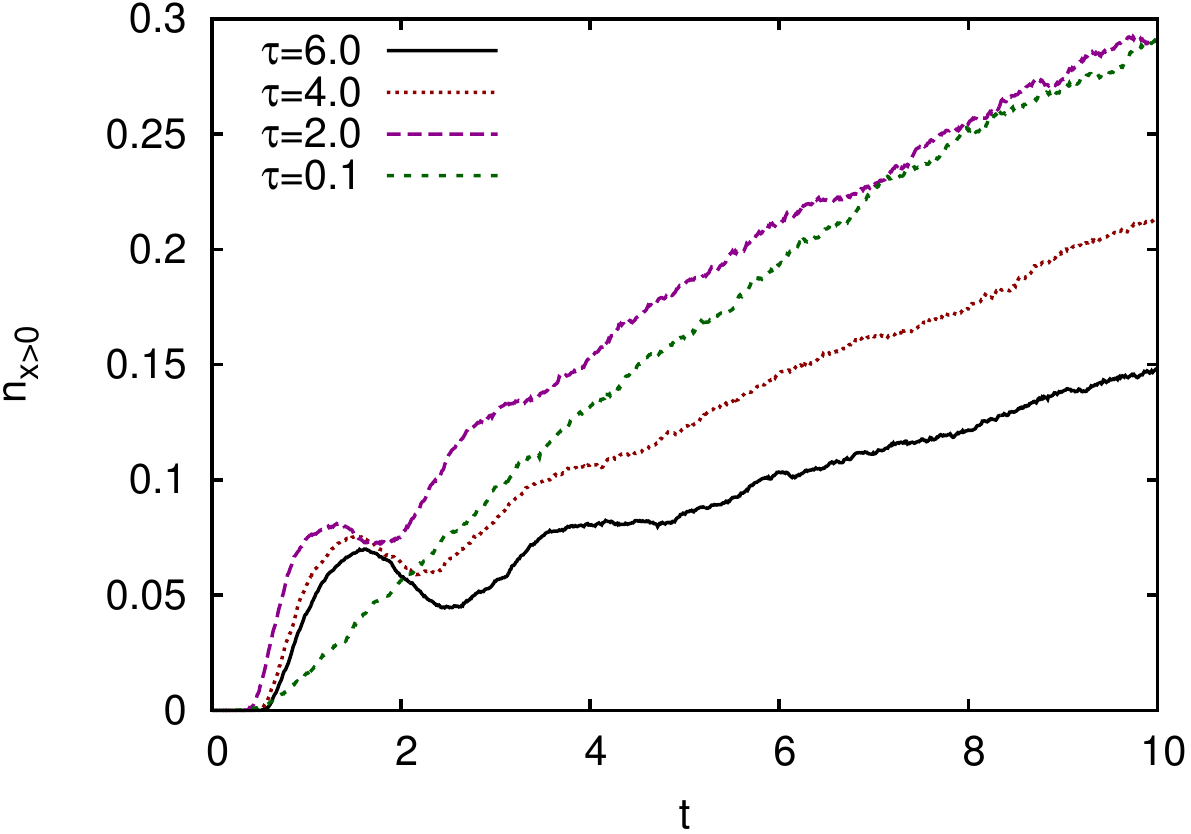}
\caption{(Color online) Relative number of particles located on the
  right side of the symmetric potential for different correlation times
  $\tau$ and two different time ranges.} \label{fig:pot2NRel}
\end{figure}

\begin{figure}[htp]
\includegraphics[width=0.9\columnwidth]{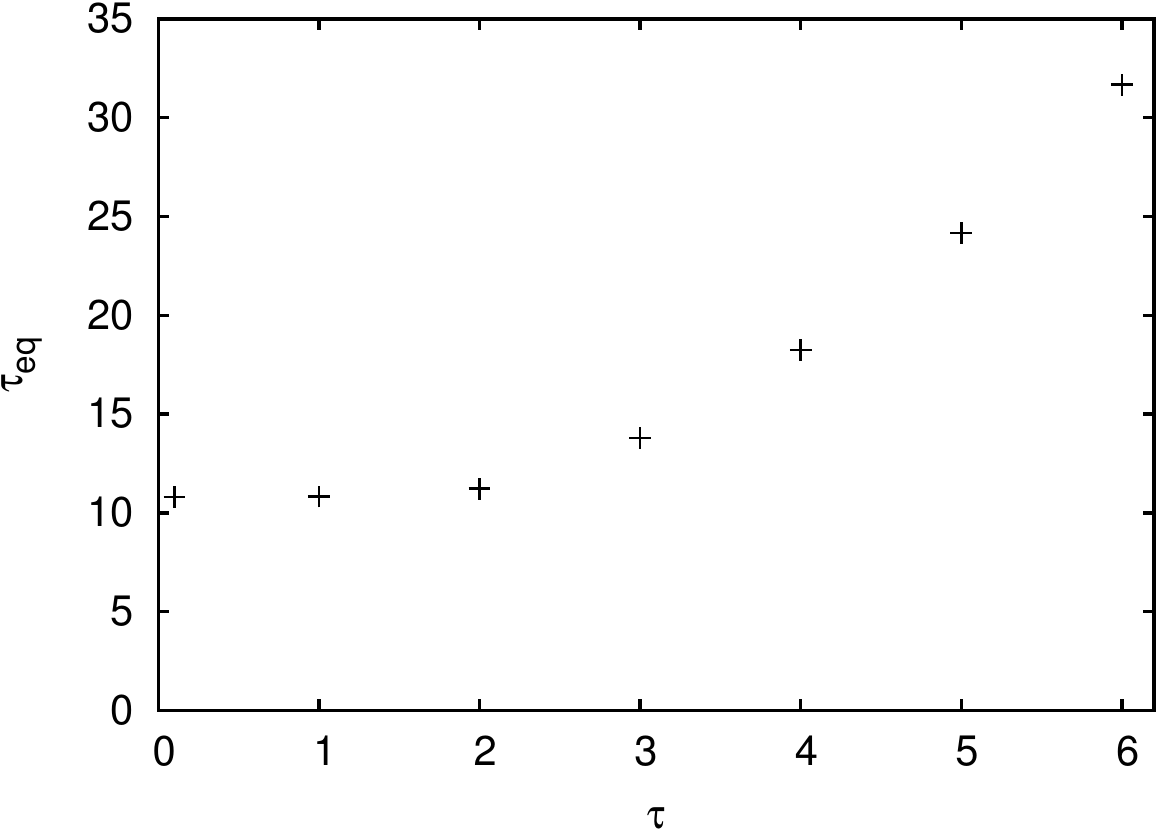}
\caption{(Color online) Characteristic time $\tau_{\mathrm{eq}}$ for the
  system to reach equilibrium in the symmetric double well potential as
  a function of the correlation time
  $\tau$.}\label{fig:equilibriumTimeN}
\end{figure}

Here, we consider a symmetric potential with $b=0$, centered around
$x=0$. For an analytic study of the diffusion over a saddle with the generalized Langevin equation,
including the exponential covariance function $C_1(t)$, see \cite{Boilley:2006mw}. 
All particles are initially located in the left potential
minimum, and the initial conditions are (see also
Fig.~\ref{fig:chapterStarterPot2})
\begin{align}
\begin{gathered}
\label{eq:pot2Init}
x_0 = x_{\mathrm{minL}}, \quad v_0 = 0, \quad D=2, \quad m=0.1, \\ T = 1,\quad
\Delta V = 1, \quad x_{\mathrm{minL}}=-x_{\mathrm{minR}}=-2 \,.
\end{gathered}
\end{align}
For the exponential covariance function $C_1(t)$ the system equilibrates
at the expected mean kinetic energy of $\frac{1}{2}T$ as illustrated
in Fig.~\ref{fig:pot2kinEnergyTau}. Let $N_{x>0}(t)$ be the number of
particles on the right side of the well and $N_{\mathrm{total}}$ the
total number of simulated particles. The relative number of particles on
the right side is then given by
\begin{equation*}
n_{x>0}(t) = N_{x>0}(t) / N_{\mathrm{total}}
\end{equation*}
and is shown for different correlation times $\tau$ in
Fig.~\ref{fig:pot2NRel}. If the correlation time is large enough, the
first particles overcoming the well (steep rise in
Fig.~\ref{fig:pot2NRel}) are dragged back to the left (drop in
Fig.~\ref{fig:pot2NRel}) due to the memory effect. When these particles
reach the left potential well they are dragged back once more from the
left to the right such that we see a rise of $n_{x>0}(t)$ again. This
oscillation could go on for a long time if the particles were not
influenced by the random force of the heat bath over time, making them
``forget'' about their history. In the case of a large correlation time
$\tau=6$ we can vaguely observe a second drop in the number of particles on the right side. 
For larger times $t>5$, $n_{x>0}(t)$ follows an exponential growth of the form
\begin{equation*}
n_{x>0}(t) = 0.5 - B\, \exp\left(-\frac{t}{\tau_{\mathrm{eq}}}\right)\,,
\end{equation*}
where $B$ and $\tau_{\mathrm{eq}}$ are two fit parameters, and
$\tau_{\mathrm{eq}}$ describes the characteristic time of the system to
reach its equilibrium state. Fig.~\ref{fig:equilibriumTimeN} shows a
significant increase of $\tau_{\mathrm{eq}}$ as a function of the
correlation time $\tau$.
	
%%%%%%%%%%%%%%%%%%%%%%%%%%%%%%%%%%%%%%%%%%%%%%%%%%%%%%%%%%%%
\subsection{Asymmetric double well potential}
\label{sec:langevinDoubleWellAsym}
%%%%%%%%%%%%%%%%%%%%%%%%%%%%%%%%%%%%%%%%%%%%%%%%%%%%%%%%%%%%

\begin{figure}[htp]
  \includegraphics[width=0.9\columnwidth]{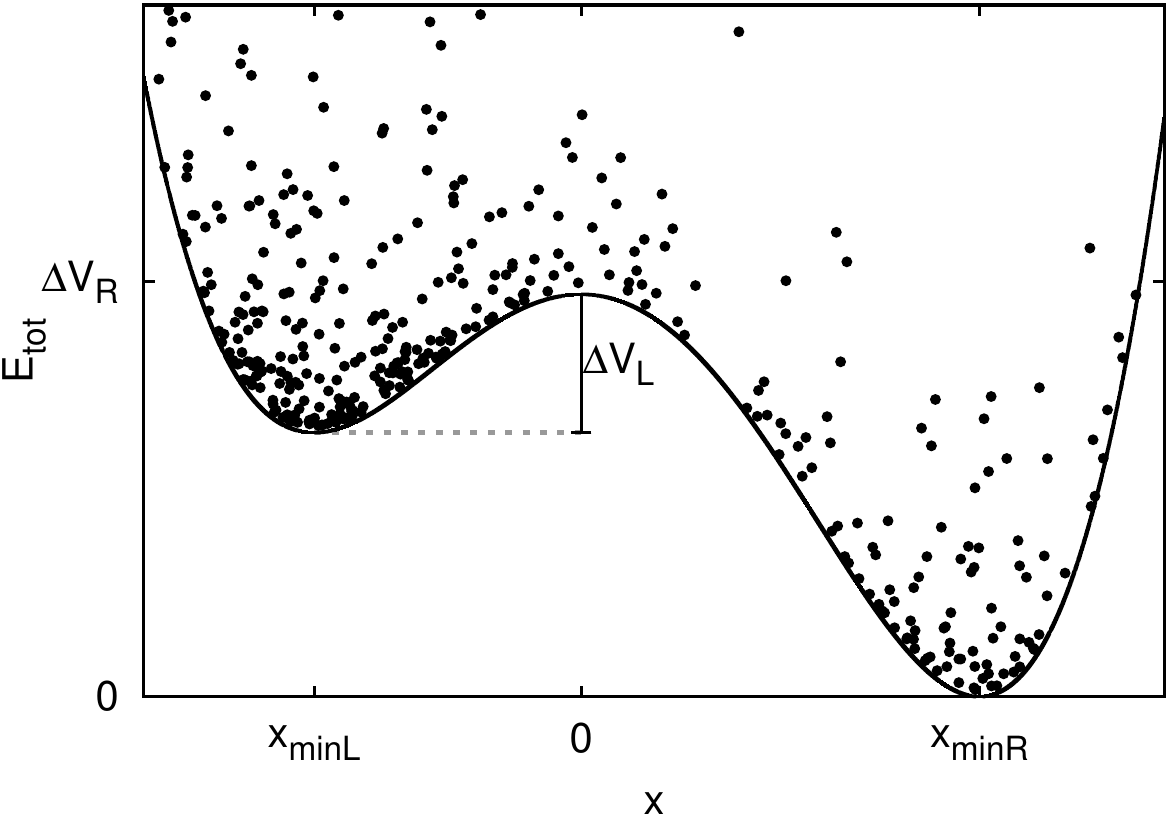}
  \caption{(Color online) 400 independent realizations of the Langevin
    equation~\eqref{eq:langevinDoubleWell} with an asymmetric double
    well potential and initial conditions as defined
    in~\eqref{eq:pot3Init}. Shown is the total energy and position of
    each particle at the time $t=5$.}
  \label{fig:chapterStarterPot3}
\end{figure}
\begin{figure}[htp]
  \includegraphics[width=0.9\columnwidth]{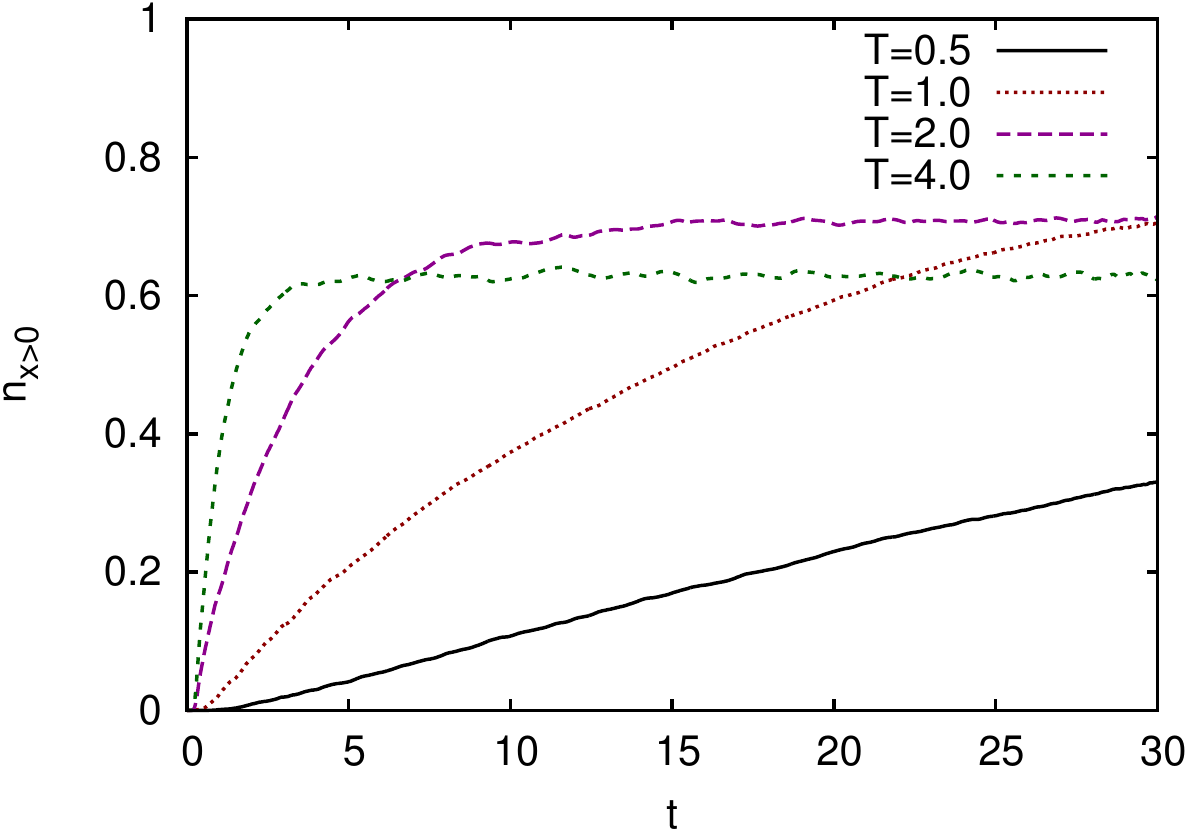}
  \caption{(Color online) Relative number of particles located on the
    right side of the potential well for different temperatures $T$ and
    correlation time $\tau=0.1$.}
  \label{fig:pot3NRelTau0_1}
\end{figure}
\begin{figure}[htp]
\includegraphics[width=0.9\columnwidth]{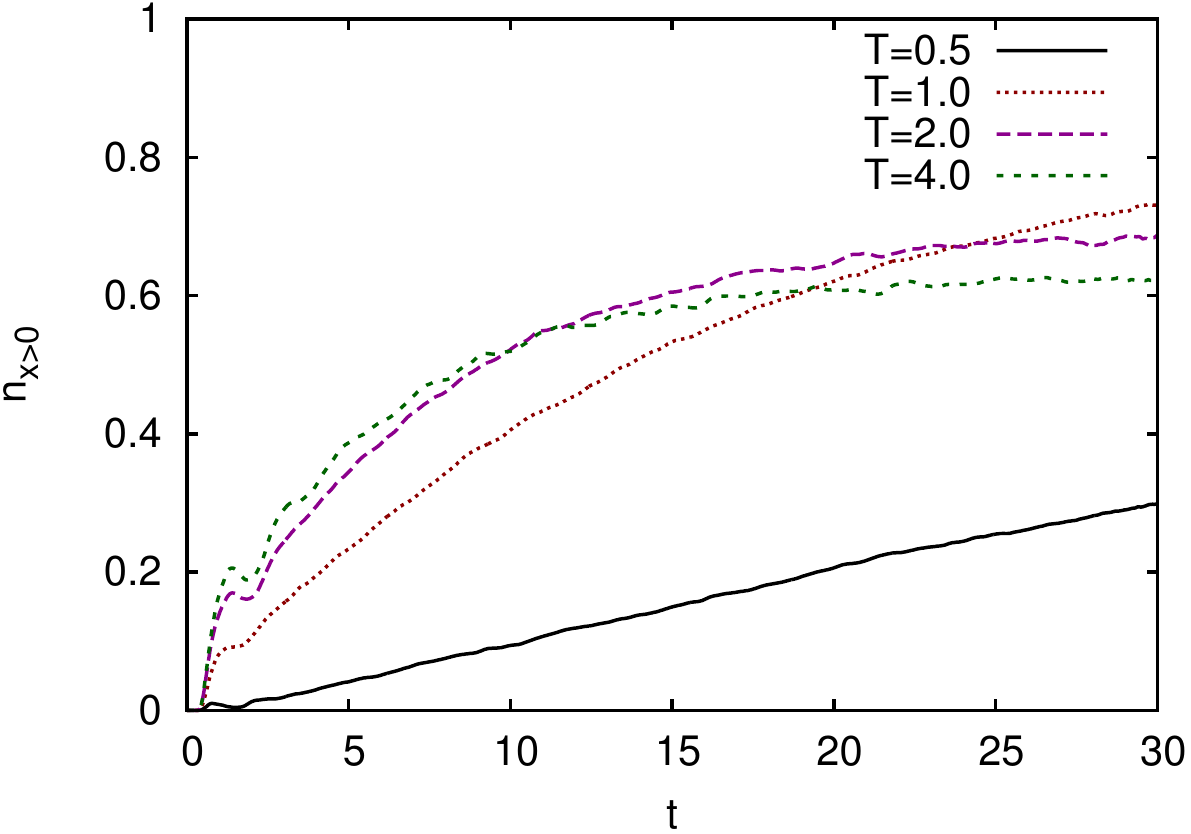}
\caption{(Color online) Same as Fig.~\ref{fig:pot3NRelTau0_1} but for
  the correlation time $\tau=2$.}
\label{fig:pot3NRelTau2_0}
\end{figure}
\begin{figure}[htp]
\includegraphics[width=0.9\columnwidth]{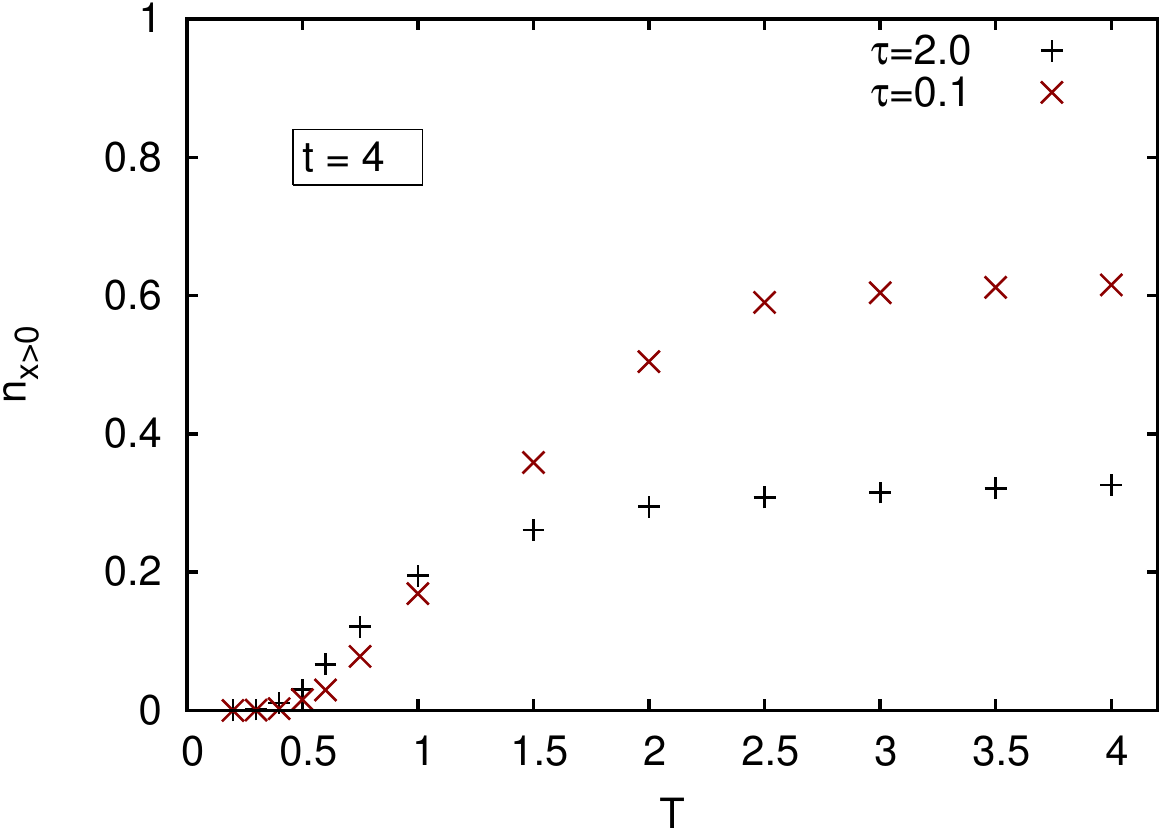}
\includegraphics[width=0.9\columnwidth]{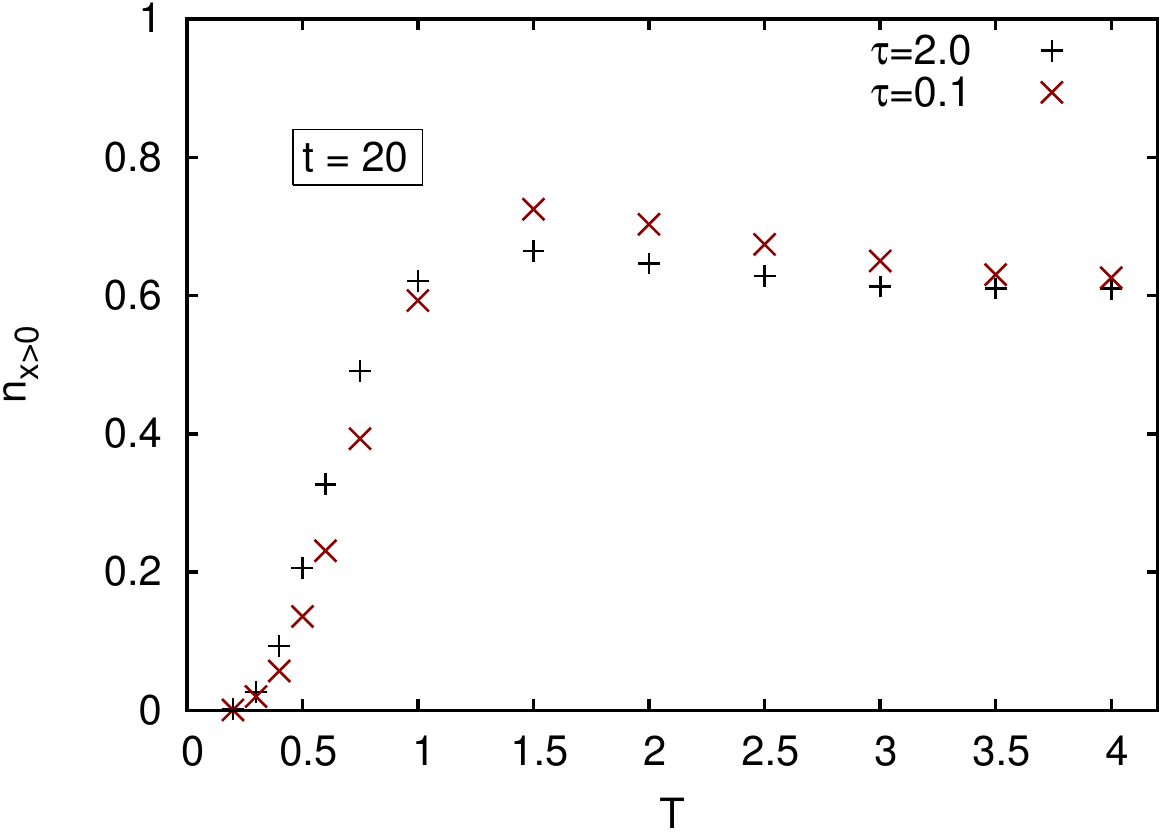}
\caption{(Color online) Relative number of particles located on the
  right side of the asymmetric potential well as a function of the
  temperature $T$ at fixed times $t=4$ (upper figure) and $t=20$ (lower
  figure). With increasing temperature more particles overcome the
  potential well in a shorter time. For very high temperatures the
  particle's energy becomes large enough to overcome the potential
  barrier as easily from the right as from the left such that the curve
  converges to $0.5$. The correlation time has a larger impact on the
  relative number of particles for smaller times $t$.}
\label{fig:pot3NRel}
\end{figure}

If $b \neq 0$, the double-well potential becomes asymmetric as shown in
Fig.~\ref{fig:chapterStarterPot3} and can be applied for instance to
describe the case of heavy-ion fusion.  For an analysis in the
white-noise limit see \cite{ABE1999}. The initial conditions for the
following simulations are
\begin{align}
\begin{gathered}
\label{eq:pot3Init}
x_0=x_{\mathrm{minL}},\quad v_0=0,\quad D=4, \quad m=0.1, \\
(\Delta V)_{\mathrm{L}} = 1, \quad (\Delta V)_{\mathrm{R}} \approx 2.90, \\
\quad x_{\mathrm{minL}} \approx -1.83, \quad x_{\mathrm{minR}} \approx
2.73\,.
\end{gathered}
\end{align}
With rising temperature some particles fall into the deeper right
potential minimum and remain there. For even higher temperatures it is
possible that these particles overcome the potential barrier
$(\Delta V)_{\mathrm{R}}$ from the right. Fig.~\ref{fig:pot3NRelTau0_1}
and \ref{fig:pot3NRelTau2_0} show the development of $n_{x>0}(t)$ for
$\tau=0.1$, $\tau=2$ and different values of the temperature. We observe
that for very high temperatures the equilibrated relative number of
right particles lowers since more particles can overcome the right
potential barrier $(\Delta V)_{\mathrm{R}}$
backwards. Fig.~\ref{fig:pot3NRel} shows this development of
$n_{x>0}(t)$ for different temperatures at two time values $t=4$ and
$t=20$. The effect of the correlation time $\tau$ becomes smaller with
increasing $t$.

%

%%%%%%%%%%%%%%%%%%%%%%%%%%%%%%%%%%%%%%%%%%%%%%%%%%%%%%%%%%%%
\section{Conclusion and outlook}
\label{sec:conclusion}
%%%%%%%%%%%%%%%%%%%%%%%%%%%%%%%%%%%%%%%%%%%%%%%%%%%%%%%%%%%%

In this paper we presented an efficient method to simulate stationary
Gaussian noise for an arbitrary covariance function and applied this
procedure to the simulation of the generalized Langevin equation with
and without external potentials leading to memory effects due to the
time correlation of the noise.

In absence of an external potential the memory effect with sufficiently
large correlation times realized by different covariance functions
manifests itself in ``plasmon'' oscillations. In the presence of a
harmonic potential we are able to solve the Langevin equation in parts
analytically and have found an effective particle oscillation frequency
composed of an oscillation due to the harmonic potential and an
oscillation due to the memory effect. Finally, we presented our
numerical results for the simulation of the generalized Langevin
equation with a symmetric and an asymmetric double well potential. Here,
we emphasize that the correlation of the noise plays an important role
in the behavior of the observed particles, leading to memory effects
that lead to a delay of the relaxation of quantities like the particle
distribution to their equilibrium values.

In this work we mainly focused on the presentation of the generalized
Langevin equation including the exponential covariance function
$C_1(t)$. While the Gaussian covariance function $C_2(t)$ also leads
to oscillations, the frequencies are different. This motivates for a
further study on the impact of different covariance functions.

Another promising investigation with the present method is the question
of the diffusion rate (Kramers rate) over a potential barrier
\cite{Haenggi:1990rmp}, which we also postpone to a future publication.

For a more general study of the generalized Langevin equation in three
dimensions, with particle-particle interaction, and a memory kernel not
only depending on $t-t'$ but on $t$ and $t'$ separately, see
\cite{Kantorovich:2008,STELLA2014}.

\begin{acknowledgments}
  We thank Eduardo Fraga for valuable discussions. We thank the
  anonymous referee for pointing us to the interesting paper
  \cite{Boilley:2006mw} which lead us to verify our algorithms on the
  test case for diffusion of a non-Markovian Brownian particle over a
  barrier. A.\ M.\ acknowledges financial support from the Helmholtz
  Research School for Quark Matter Studies (H-QM) and HIC for FAIR. H.\
  v.\ H.\ has been supported by the Deutsche Forschungsgemeinschaft
  (DFG) under grant number GR 1536/8-1.
\end{acknowledgments}

\appendix*
\section{Algorithm for colored noise}
\label{sec:appendix}

Here, we give explicitly a basic numerical algorithm for generating a
sequence of colored noise in accordance with our method, which is
described in \ref{sec:noise}.
\begin{enumerate}
\item Define a sufficiently large time interval $[-\Delta, \Delta]$,
  such that all relevant functions (see below for $C(t), G(t)$) become
  negligibly small outside the interval. Using $(2M+1)$ equidistant grid
  points on this interval lead then to the following discretized time
  values and frequency modes:
\begin{equation}
\begin{split}
  \Delta t=\frac{\Delta}{M}\,\,&\Rightarrow\,\, t_m=m\cdot\Delta t\\
  \Delta\omega = \frac{2\pi}{\left(2M+1\right)\Delta
    t}\,\,&\Rightarrow\,\,\omega_n = \frac{2\pi
    n}{\left(2M+1\right)\Delta t}
\end{split}
\end{equation}
with $m, n\in\{-M, ..., 0, ..., M\}$
\label{item:1}
\item Take the Fourier transform
  $S_\xi(\omega) = \mathcal{F}[C](\omega)$ of the desired covariance
  function $C(t)$ such as given in \eqref{eq:covFunc} on the discrete
  set of $t$-values:
\begin{equation}
 S_\xi\left(\omega_n\right) = \Delta t\sum_{m=-M}^M
 C\left(t_m\right)\exp(\ii\omega_n t_m) \,.
\end{equation}
\label{item:2}
\item Take the inverse Fourier transform of $\sqrt{S_\xi(\omega)}$ on
  the discrete set of $\omega$-values:
\begin{equation}
  G\left(t_m\right)=\frac{\Delta\omega}{2\pi}\sum_{n=-M}^M\sqrt{S_\xi\left(\omega_n\right)}
  \exp (-i\omega_n t_m)\,.
\end{equation}
\item Generate a sequence of white noise $\xi_w(t)$ on the time interval
  $[-\Delta, \Delta + T]$:
\begin{equation}
 \xi_w(t_i) =\frac{\bar a_i}{\sqrt{\Delta t}}\,,
\end{equation}
where $T\geq 0$ defines a time interval for colored noise. The variable
$\bar a_i$ is a standard normally distributed random number.
\item Take the convolution of $G(t)$ with $\xi_w(t)$ on the time interval $[-\Delta, \Delta]$ to generate a sequence of colored noise:
\begin{equation}
 \xi(t_j)=\Delta t\sum_{m=-M}^M G\left(t_m\right)\xi_w\left(t_j+t_m\right)
\end{equation}
with $t_j=j\cdot\Delta t$ for $j\in\{0,..,N\}$ denoting the time points
on $[0, T]$.

We note that for every sequence of colored noise $\xi(t)$ the white
noise $\xi_w(t)$ has to be generated independently.

\label{item:3}
\end{enumerate}
\newpage
\bibliography{langevinColouredNoise}
\end{document}